\documentclass[a4paper, 11pt]{article}
\usepackage{indentfirst}
\pdfoutput=1

\usepackage{jinstpub}

\title{Cherenkov and Scintillation Light Separation Using Wavelength in LAB Based Liquid Scintillator} 
\author[a,1]{Tanner Kaptanoglu,\note{Corresponding Author.}}
\author[a]{Meng Luo,}
\author[a]{Josh Klein}
\affiliation[a]{University of Pennsylvania, \\ 209 South 33rd Street Philadelpha PA 19104, USA}
\emailAdd{tannerk@hep.upenn.edu}

\abstract{Linear alkyl benzene (LAB) has in recent years been used as a solvent for PPO in large-scale scintillation detectors, like Daya Bay and SNO+. The combination has several nice properties, including high light yield, good materials compatibility, and excellent pulse shape discrimination. As charged particles move through the LAB+PPO, both Cherenkov and scintillation light are created. Separating Cherenkov from scintillation light would allow a broad range of physics in future large-scale detectors like THEIA, by allowing direction reconstruction with Cherenkov light while retaining the high light yield and good particle ID of a scintillator detector. In this paper, we examine the discrimination of Cherenkov and scintillation light using a set of bandpass and dichroic filters. In principle, Cherenkov light emission extends longer in wavelength than the PPO scintillation spectrum, allowing for exclusive identification. We find that by selecting wavelengths above 450~nm the Cherenkov light can be clearly separated from the scintillation light.}


\begin{document}

\maketitle

\section{Introduction}

Several large-scale neutrino experiments use liquid scintillator targets to achieve a high light yield and excellent pulse-shape discrimination. Some of these experiments, as well as future experiments, rely on Linear alkyl benzene (LAB) as the primary solvent in a scintillator cocktail~\cite{snoplus, juno, dayabay2} that often includes 2,5-Diphenyloxazole (PPO) as a primary fluor.  In a large-scale detector like SNO+ or Daya Bay, the wavelength profile of the PPO (and any additional secondary fluors) plays an important role in the detector's overall light yield, because non-trivial amounts of scattering and absorption affect the number and spectrum of photons reaching the photon detectors (photomultiplier tubes (PMTs) or other devices). In addition to PPO, a fluor called p-terphenyl (PTP) is investigated in this paper as well.

Future large-scale scintillation experiments like THEIA~\cite{theia} plan to detect both Cherenkov and scintillation light as a way of providing a very broad range of physics with a single detector. Cherenkov light allows direction reconstruction which can aid in identifying solar neutrino events~\cite{rbonvgdog}, identification of neutrinoless beta decay candidates against the solar neutrino background~\cite{biller}, or discrimination of high-energy $\nu_e$ events from $\pi^0$s, which is  important for studying long-baseline neutrino oscillations.  The scintillation light provides a high light yield that is critical for good energy and position reconstruction. The time profile of the scintillation light is also important, because it affects position reconstruction and provides ways of discriminating particles like $\beta$s from $\alpha$s. The emission time profile has already been well characterized on the bench-top by several experiments including SNO+, Daya Bay, LENA, and Borexino~\cite{dayabay, queens, lena, ranucci}. Separation of Cherenkov and scintillation light using the difference in timing and the directionality of the Cherenkov light was demonstrated very nicely in ~\cite{chsp, bh, berkeley}.

Another approach for discriminating Cherenkov from scintillation light is to look in different wavelength bands. Cherenkov light is much broader, peaking in the UV and falling off like ${\lambda^{-2}}$, while scintillation light typically has a narrow-band spectrum.  Thus, by selecting long wavelength light, either using filters or red-sensitive photon detectors, we anticipate an ability to detect some Cherenkov photons without contamination from scintillation light~\cite{chsp}. 

A critical requirement for this is that the scintillator emission spectrum is very small at long wavelengths, so that the Cherenkov light is not hidden. LAB+PPO would therefore seem like an ideal cocktail: its spectrum spans a relatively narrow band of wavelengths, ranging from about 350 to 450~nm. We find in this paper that above about 450~nm the Cherenkov light can be clearly identified over the scintillation light. We perform fits to the Cherenkov and scintillation spectra that show around 490~nm one can select a 95\% pure sample of Cherenkov light using a prompt-time selection. 

\section{Cherenkov and Scintillation Separation Using Bandpass Filters}

\subsection{Experimental Setup}\label{sec:experimental}

\begin{figure}
    \centering
    \includegraphics[trim = 2.0cm 12.5cm 0.5cm 9cm, clip=true, scale=0.6]{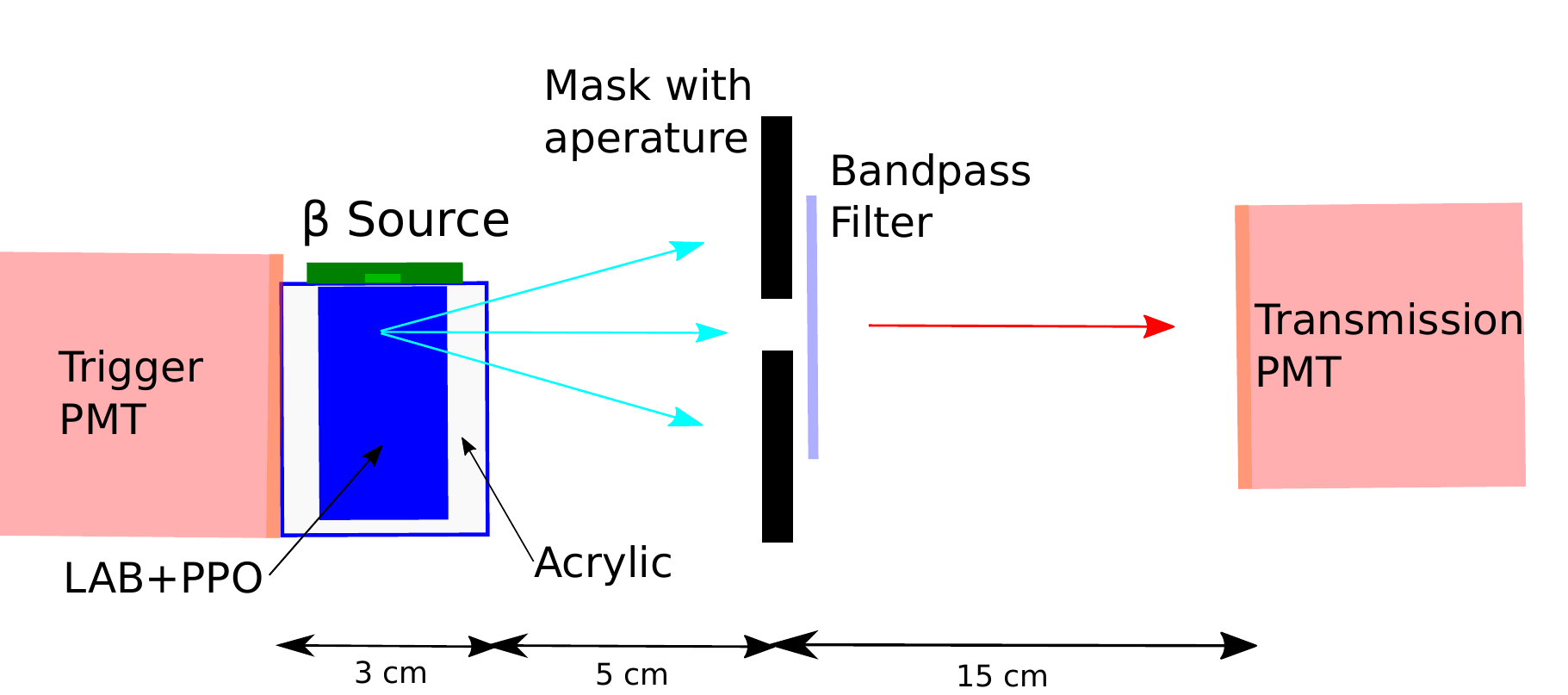}
    \caption{A schematic of the experimental setup, showing the $\beta$ source deployed above the LAB+PPO target and the locations of the mask, PMTs, and bandpass filter. The colored lines indicate example optical photon paths before and after the filter. The aperture is 1 cm in diameter.}
    \label{fig:schematic}
\end{figure}

A diagram of the experimental setup is shown in Figure \ref{fig:schematic}. The setup consists of a $^{90}$Sr $\beta^{-}$ source deployed above a hollowed out UV-transparent acrylic block. The block is a 3 x 3 x 3.5 cm cube with a machined out cylindrical volume that is 2~cm in diameter. That volume is filled with LAB+PPO scintillator. The PPO is added at a concentration of 2~g/L to the LAB. Dissolved oxygen, which has previously been shown to impact the time profile of the scintillation light~\cite{queens}, was not removed from the scintillator.  The majority of the effect is the removal of some quenching for the late-time scintillation light, but the first time-constant is not significantly changed, which is the primary concern for this paper. 

The $^{90}$Sr undergoes a 0.546~MeV $\beta^{-}$ decay to $^{90}$Y with a half life of 29.1 years. The $^{90}$Y undergoes a 2.28~MeV $\beta^{-}$ decay to $^{90}$Zr with a half-life of 64 hours. The $\beta^{-}$ particles enter the scintillator and create isotropic scintillation light. In addition, the $\beta^{-}$ particles create Cherenkov light in the scintillator. At these low energies the $\beta^{-}$ particles often undergo several scatters such that the Cherenkov light can be created traveling in any direction. The light yield of the scintillation light is about 10,000 photons per MeV~\cite{bh}, compared to around 300 photons per MeV for Cherenkov light in the wavelength band detectable by photomultiplier tubes. 

\begin{figure}[ht!]
\centering 
\includegraphics[scale=0.6]{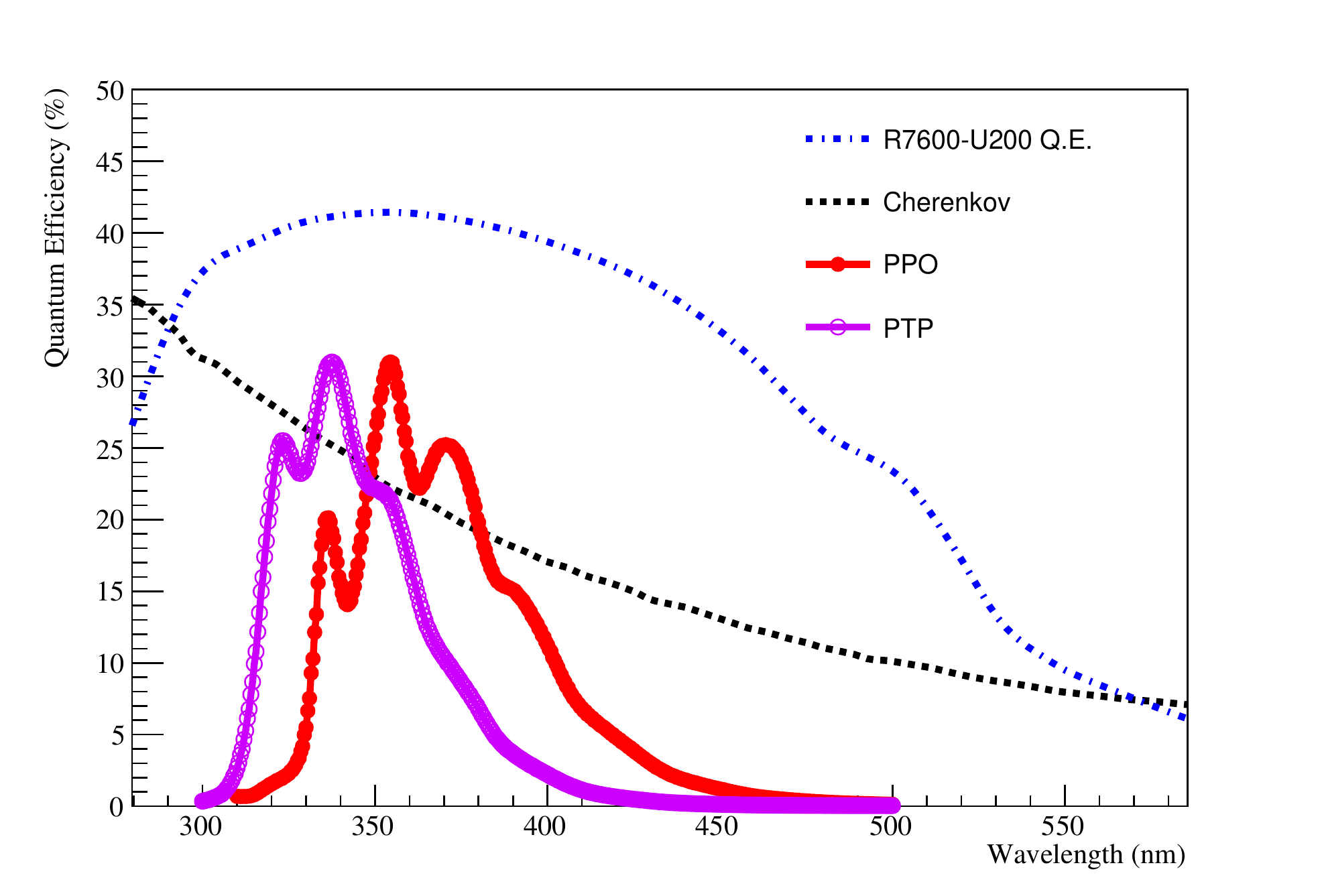}
\caption{The wavelength spectrum of Cherenkov light, the PPO emission, the PTP emission, and the R7600-U200 PMT quantum efficiency (Q.E.). The PPO and PTP emission spectra are provided by the PhotochemCAD database \cite{photochem}. The PPO, PTP, and Cherenkov spectra are scaled arbitrarily and show shape only. The Q.E. curve is provided by Hamamatsu~\cite{hamamatsu} and peaks above 40\%.}
\label{fig:spectra}
\end{figure}

A Hamamatsu super-bialkali R7600-U200 PMT is optically coupled to the acrylic block using Saint-Gobain BC-630 optical grease. The PMT acts as a high efficiency fast trigger and provides the time-zero. The efficiency curve of the PMT, shown in Figure \ref{fig:spectra}, spans a broad range of wavelengths and can thus be used for all of the measurements described in this paper. On the other side of the source, 5 cm from the acrylic block, is a mask with a small aperture, 1 cm in diameter. The center of the aperture is 3 cm above the base of the acrylic block. Directly behind the aperture we place a series of bandpass filters that absorbs light outside a small band of wavelengths. The size of the aperture ensures that the transmission PMT detects primarily single photons. That light is viewed by a second R7600-U200 PMT, placed 15 cm from the mask, which is used to measure the transmitted light over the wavelengths passed by the filter.  This low coincidence rate technique follows the single photon counting method described in~\cite{spc}. Both PMTs are operated at -800~V where their gains are approximately $5 \times 10^{6}$.

Our data acquisition system includes a Lecroy WaveRunner 606Zi 600 MHz oscilloscope, which is used to digitize the signals from both PMTs. Waveforms are sampled every 100~ps for 500~ns. The oscilloscope has an 8-bit ADC with a variable dynamic range, which allows for roughly 100 $\mu$V resolution. The LeCrunch software~\cite{lecrunch} is used to read out the data from the oscilloscope over Ethernet as well as format the data into custom HDF5 files.

The bandpass filters that are used in this measurement are shown in Table \ref{tab:filters}. A set of filters are chosen to span across the emission spectrum of LAB+PPO. Data provided by Thorlabs and Edmund Optics~\cite{thorlabs, edmund} indicate that the transmission is less than 0.01\% outside of the pass band of the filters. In all cases, the filter is placed normal to the light coming from the aperture, so that the average incident angle at the filter is 0$^{\circ}$. 

\begin{table}[ht!]
    \centering
    \begin{tabular}{|l|l|l|} \hline 
         \textbf{Center (nm)} & \textbf{FWHM (nm)} & \textbf{Peak Transmission (\%)} \\ \hline \hline 
         355 & 10 & 95 \\ \hline 
         387 & 11 & 95 \\ \hline 
         405 & 10 & 96\\ \hline 
         430 & 10 & 46 \\ \hline 
         450 & 10 & 98 \\ \hline 
         470 & 10 & 53 \\ \hline 
         494 & 20 & 95 \\ \hline 
         510 & 10 & 60 \\ \hline 
         530 & 10 & 54 \\ \hline 
    \end{tabular}
    \caption{The central wavelength, FWHM, and transmission at the central wavelength for each of the bandpass filters. The tolerances on the central wavelengths of the filters are less than or equal to 3~nm. Data is provided by Thorlabs and Edmund Optics~\cite{thorlabs, edmund}.}
    \label{tab:filters}
\end{table}

\subsection{Data Analysis}\label{sec:analysis}

Offline analysis code is used to find coincidences between the trigger and transmission PMTs. A software-based constant fraction discriminator is used to find the time difference between the trigger PMT and the transmission PMT signals. This is done by scanning the digitized waveform of the transmission PMT looking for a threshold crossing that is 3 times larger than the width of the electronics noise. For each threshold crossing, a 15~ns window is chosen and the number of consecutive samples above threshold are counted. If the threshold is crossed for longer than 3~ns, the analysis flags the threshold crossing as associated with a true PMT pulse, rather than a spike in the electronics noise. The peak of the PMT pulse is identified, and the sample associated with the 20\% peak-height crossing is found. This sample is taken as the time of the PMT pulse for the transmission PMT. The same is done for the trigger PMT, where it is easy to identify the large signal in every event. The analysis allows for multiple threshold crossings, and thus multiple PMT pulses, in the same waveform, although this ends up having a negligible impact on the analysis.

A very low coincidence rate of less than 3\% ensures that the transmission PMT is detecting primarily single photons. The $\Delta$t between the trigger and transmission PMT crossing is histogrammed. The only event selection criterion is a lower charge cut on the trigger PMT to ensure the triggered event corresponds to the $\beta^{-}$ particle traveling through the scintillator.

Another particularly nice feature of the R7600-U200 PMTs is the negligible amounts of pre-pulsing, late-pulsing, and dark-pulsing~\cite{etel}. The dark count rate of the PMTs is around 50 Hz and is uncorrelated with the trigger time. As a result, the contribution from dark counts to the $\Delta$t histogram is insignificant, which is confirmed for each data set by looking in a window before the trigger. Because the trigger rate from the $\beta^{-}$ source is approximately 1 kHz, contributions are minor from backgrounds such as cosmic muons through the scintillator volume, which was confirmed by taking data with no source.

\subsection{Results}\label{sec:results}

Data was taken for nine different bandpass filters with distinct central transmission wavelengths, shown in Table \ref{tab:filters}. The scintillation time profile is constructed for each data set using the methods described in Section \ref{sec:analysis}. Less data was necessary for the shorter wavelength bandpass filters to achieve roughly the same level of statistics as the longer wavelength filters. In general, sufficient data was taken for each filter until the statistical uncertainties were smaller than 3\% in the peak of the emission spectrum. The 3\% level ensures that the rise times of the time profiles can be clearly distinguished. Because very little scintillation light is detected when using the bandpass filters above 500~nm, the statistics are lower for those data sets. Data was also taken with the aperture masked off with black felt. This data set showed a coincident rate consistent with the dark rate of the PMTs. Additionally, a data set was taken with no bandpass filter to extract the time profile integrated across the entire PPO emission spectrum.

\begin{figure}[ht!]
\centering 
\includegraphics[scale=0.6]{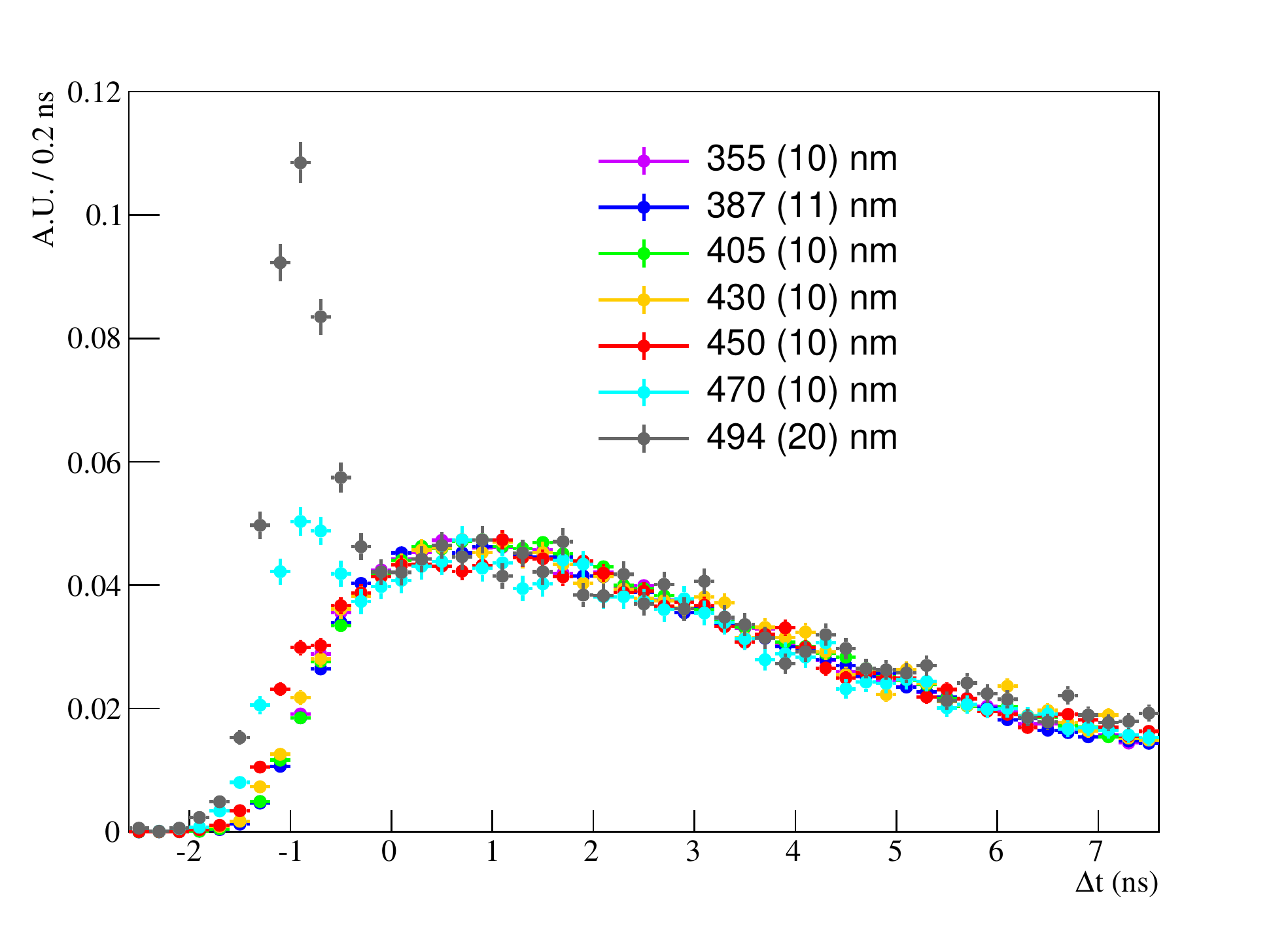}
\caption{The time emission profile, zoomed into the rise time of the LAB+PPO for different wavelength regimes. The light is selected using the bandpass filters listed in Table \ref{tab:filters}. The central wavelength and width of the bandpass filters are specified in the legends. The histograms are normalized to the peak of the scintillation light and shown in arbitrary units (A.U.). The Cherenkov light can be clearly identified at early times in the data for the filters longer than 450~nm. These histograms, as well as additional ones, are shown separately in Figure \ref{fig:timing}.}
\label{fig:selectedbandpass}
\end{figure}

\begin{figure}[ht!]
\centering 
\includegraphics[scale=0.8]{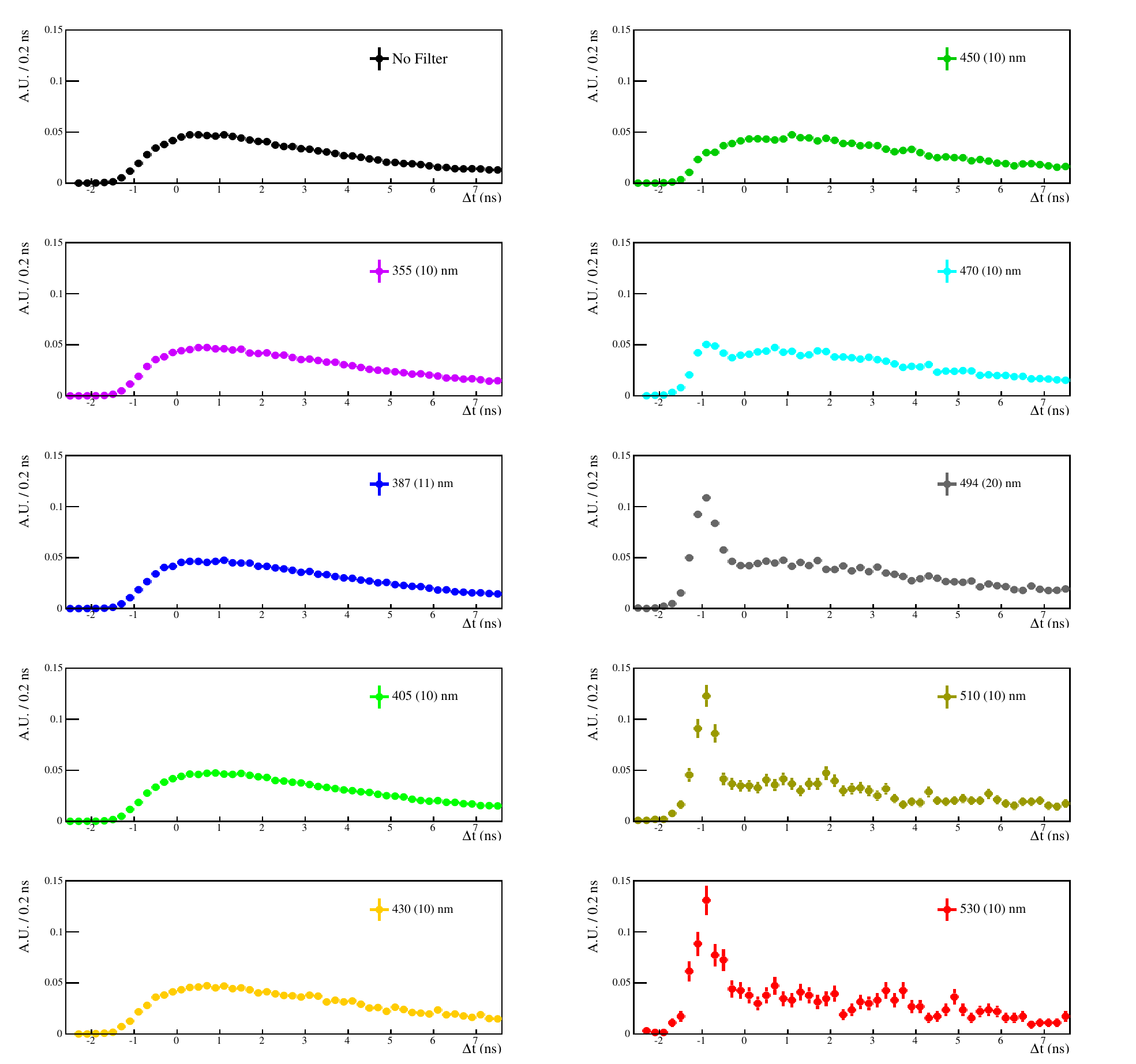}
\caption{The time emission profile for each of the bandpass filters specified in Table \ref{tab:filters}. Each of the data sets are normalized to the peak of the scintillation light. The Cherenkov light becomes clearly separated above 450~nm.}
\label{fig:timing}
\end{figure}

Figure \ref{fig:selectedbandpass} shows the scintillation emission profile for several of the data sets. The histograms are normalized to the peak of the scintillation light in order to directly compare the scintillation profile. This figure makes clear how the Cherenkov light becomes a larger fraction of the prompt light at longer wavelengths. Additionally, it shows the first time-constant of the time profiles are consistent for the scintillation light at wavelengths spanning the entire PPO emission spectrum. Figure \ref{fig:timing} shows the histograms for all of the bandpass filters in Table \ref{tab:filters}.

The spectra are fit to Equation \ref{eq:fit} using the RooFit package~\cite{roofit}. The scintillation light is fit to the sum of two decay exponentials with time constants $\tau_{1}$ and $\tau_{2}$, weights $A_{1}$ and $A_{2}$, and an exponential rise time, $\tau_{R}$. The weights $A_{1}$ and $A_{2}$ are constrained to sum to one. The PMT transit time spread (TTS) is determined offline using a Cherenkov light source to be 350 ps, which is in good agreement with the Hamamatsu datasheet. Additionally, $f_{PMT}$ accounts for the TTS of both the trigger and transmission PMTs. That distribution is convolved with the scintillation profile in the fit. An offset $t'$ allows for cable delays and other arbitrary time offsets. The second component of the model accounts for the Cherenkov light, which is modeled simply as $f_{PMT}$ because the Cherenkov light is emitted promptly in comparison to the TTS of the PMTs. The scintillation and Cherenkov light are weighted appropriately by $C$. 

\begin{equation}\label{eq:fit}
    F =  C \times f_{PMT}(t - t') + (1 - C) \times \sum_{i=1}^{2} \frac{A_{i} \times (e^{-t/\tau_{i}} - e^{-t/\tau_{R}})}{(\tau_{i} - \tau_{R})} * f_{PMT}(t - t')
\end{equation}

\begin{figure}[ht!]
\centering 
\includegraphics[width=0.495\textwidth]{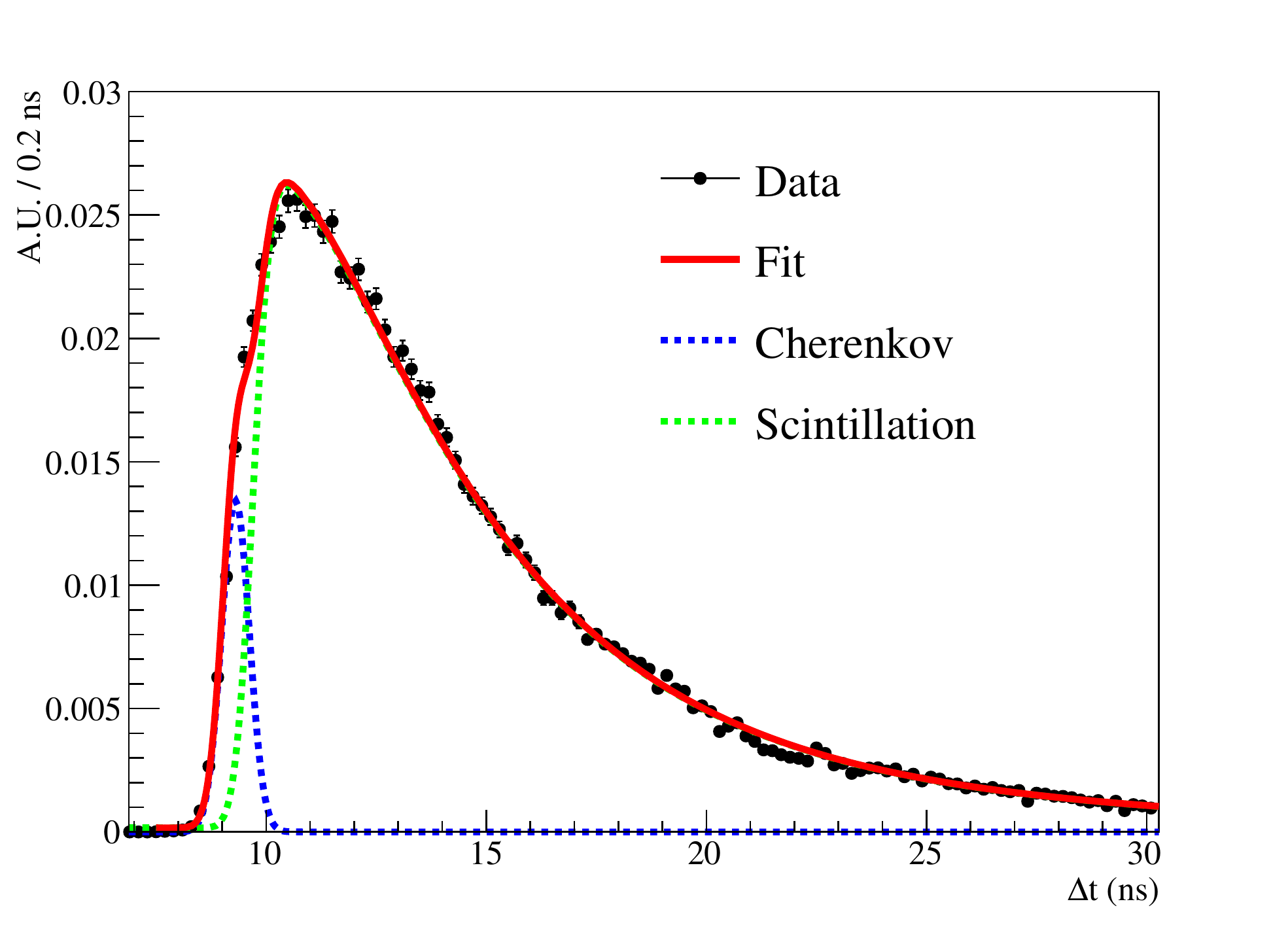}
\includegraphics[width=0.495\textwidth]{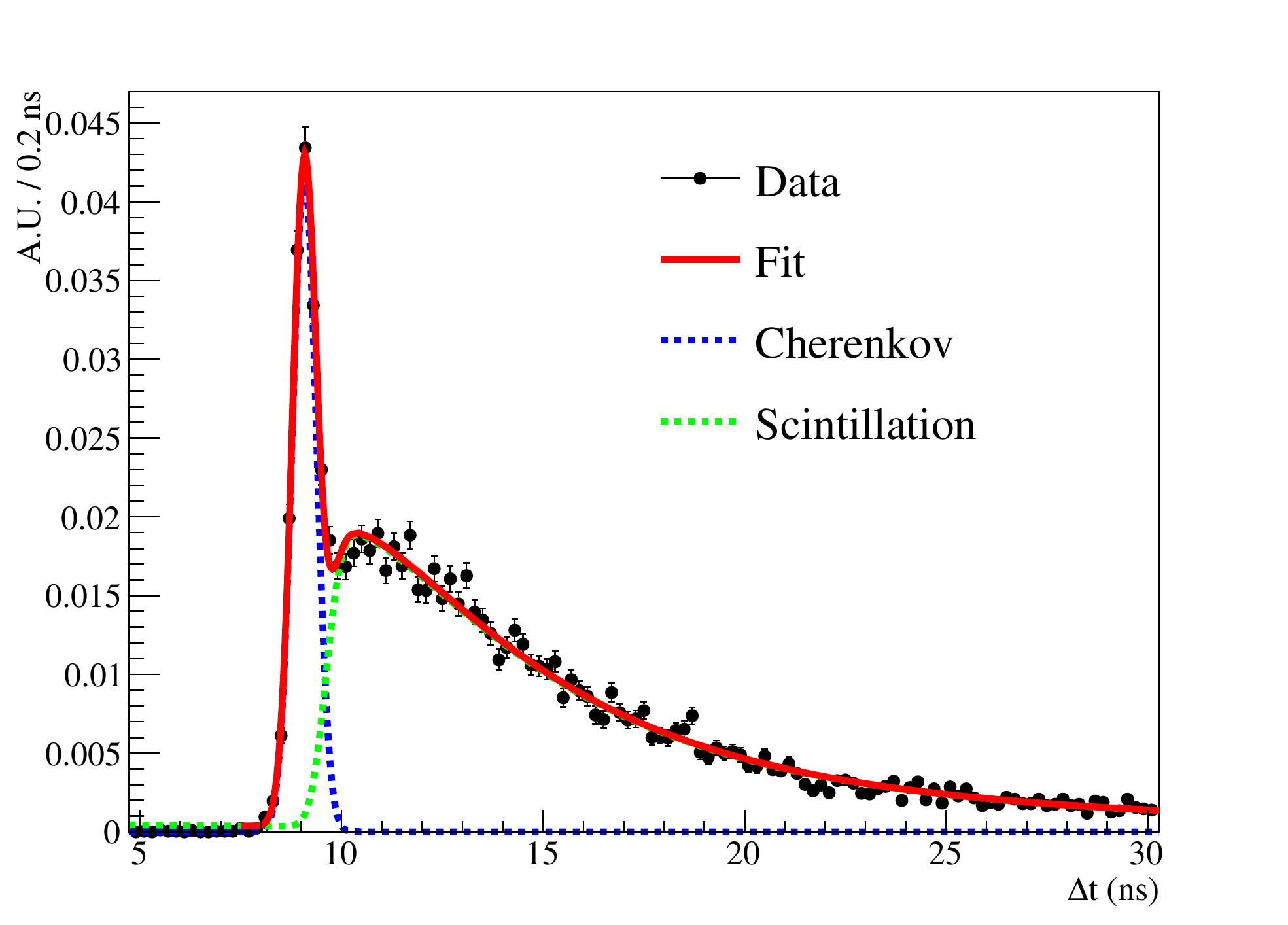}
\caption{The full fit for the 355~nm bandpass filter data (left) and 494~nm bandpass filter data (right) is shown in red. The Cherenkov and scintillation components of the fit are shown explicitly. Note the arbitrary $\Delta$t offset from 0~ns, mostly due to cable delays, was not removed for this plot, as it was for Figures \ref{fig:selectedbandpass} and \ref{fig:timing}.}
\label{fig:fit1}
\end{figure}

The full scintillation spectrum is typically fit with three or four exponential decay time constants. Our primary goal here, however, is to identify the Cherenkov light, and thus only the first 30 ns of the spectrum are fit. Neither the length of the waveform nor the length of data taking provides for accurate measurement of the full scintillation spectrum, which has already been measured by several experiments. Figure \ref{fig:fit1} shows two examples of the fit with the Cherenkov and scintillation components separated. 

One can select a relatively pure sample of Cherenkov light by integrating the time profile in a prompt window. Equation \ref{eq:acceptance} defines the purity, $P$, of the Cherenkov light in a prompt window, where $F_{C}$ is the Cherenkov component in the fit. The upper and lower bounds of the integral are taken to span the Cherenkov component (Figure \ref{fig:fit1}) and are arbitrarily offset from zero based on cable delays and other time offsets.  

\begin{equation}\label{eq:acceptance}
    P = \int \limits_{8.0}^{9.5} \frac{F_{C}}{F} dt
\end{equation}

Table \ref{tab:fitresults} shows the full fit results for the ten data sets. The time constants and their relative fractions ($\tau_{1}$, $\tau_{2}$, and $A_{1}$) are consistent across a wide range of wavelengths. This suggests that the scintillation light in each wavelength regime is being produced via the same mechanism. The purity, $P$, of the Cherenkov light selection is high even for the short wavelength bandpass filters due to the excellent timing characteristics of the PMTs used. The fits to the data for filters above 450~nm show the Cherenkov light can be selected with a purity of over 90\%. This is apparent in Figure \ref{fig:fit1} (right) where the fit to the 494~nm bandpass filter shows that the prompt window is dominated by the Cherenkov component. 

\begin{table}[ht!]
    \centering
    \begin{tabular}{|c|c|c|c|c|c|c|} \hline 
         \textbf{Filter} & \textbf{$\boldsymbol{A_{1}}$ (\%)} & \textbf{$\boldsymbol{\tau_{R}}$ (ns)} & \textbf{$\boldsymbol{\tau_{1}}$ (ns)} & \textbf{$\boldsymbol{\tau_{2}}$ (ns)} & \textbf{$\boldsymbol{1 - C}$ (\%)} & \textbf{P (\%)} \\ \hline \hline  
         None & 68 $\pm$ 3 & 1.6 $\pm$ 0.1 & 3.6 $\pm$ 0.1 & 11.2 $\pm$ 1.2 & 94 $\pm$ 2 & 73 $\pm$ 4\\ \hline 
         355 & 71 $\pm$ 4 & 1.7 $\pm$ 0.1 & 3.6 $\pm$ 0.1 & 10.5 $\pm$ 1.8 & 94 $\pm$ 2 & 70 $\pm$ 4 \\ \hline 
         387 & 73 $\pm$ 4 & 1.6 $\pm$ 0.1 & 3.7 $\pm$ 0.1 & 10.8 $\pm$ 1.8 & 95 $\pm$ 2 & 69 $\pm$ 4\\ \hline 
         405 & 68 $\pm$ 3 & 1.7 $\pm$ 0.1 & 3.6 $\pm$ 0.1 & 10.1 $\pm$ 2.0 & 96 $\pm$ 2 & 68 $\pm$ 4 \\ \hline 
         430 & 70 $\pm$ 3 & 1.7 $\pm$ 0.1 & 3.6 $\pm$ 0.1 & 11.4 $\pm$ 1.8 & 94 $\pm$ 2 & 78 $\pm$ 4 \\ \hline 
         450 & 68 $\pm$ 3 & 1.7 $\pm$ 0.1 & 4.0 $\pm$ 0.2 & 11.6 $\pm$ 2.2 & 93 $\pm$ 2 & 76 $\pm$ 4 \\ \hline 
         470 & 68 $\pm$ 4 & 1.7 $\pm$ 0.2 & 4.0 $\pm$ 0.2 & 11.9 $\pm$ 2.4 & 89 $\pm$ 2 & 90 $\pm$ 4 \\ \hline 
         494 & 68 $\pm$ 3 & 1.7 $\pm$ 0.1 & 4.0 $\pm$ 0.2 & 12.0 $\pm$ 2.0 & 82 $\pm$ 2 & 95 $\pm$ 4 \\ \hline 
         510 & 68 $\pm$ 4 & 1.7 $\pm$ 0.3 & 4.0 $\pm$ 0.3 & 12.0 $\pm$ 3.2 & 80 $\pm$ 4 & 98 $\pm$ $^{2}_{5}$ \\ \hline 
         530 & 68 $\pm$ 6 & 1.8 $\pm$ 0.4 & 4.0 $\pm$ 0.4 & 11.9 $\pm$ 4.2 & 77 $\pm$ 7 & 98 $\pm$ $^{2}_{7}$ \\ \hline 
    \end{tabular}
    \caption{The fit results for each of the bandpass filters. The fit parameters are defined in Equation \ref{eq:fit}. P is the purity of the Cherenkov light selected in a prompt window and is defined in Equation \ref{eq:acceptance}.}
    \label{tab:fitresults}
\end{table}

The amount of scintillation light detected for each bandpass filter can be compared to the expectation from the PPO emission spectrum. In order to this, the PPO emission spectrum output from our set-up must be measured. The emission spectrum of the LAB+PPO is measured using a UV-Vis Ocean Optics Spectrometer by directly exciting the scintillator with a 335~nm LED. The measurement is made by placing the spectrometer at the same relative location as the aperture to probe the spectrum of light that the transmission PMT will view. The scintillator for this measurement is held in the same acrylic block. The PPO absorbs strongly on its own emission spectrum, which causes the lack of the short wavelength peak when comparing to Figure \ref{fig:spectra}. The self-absorption of PPO is described in more detail in~\cite{selfabsorption}.

\begin{figure}[ht!]
\centering 
\includegraphics[scale=0.6]{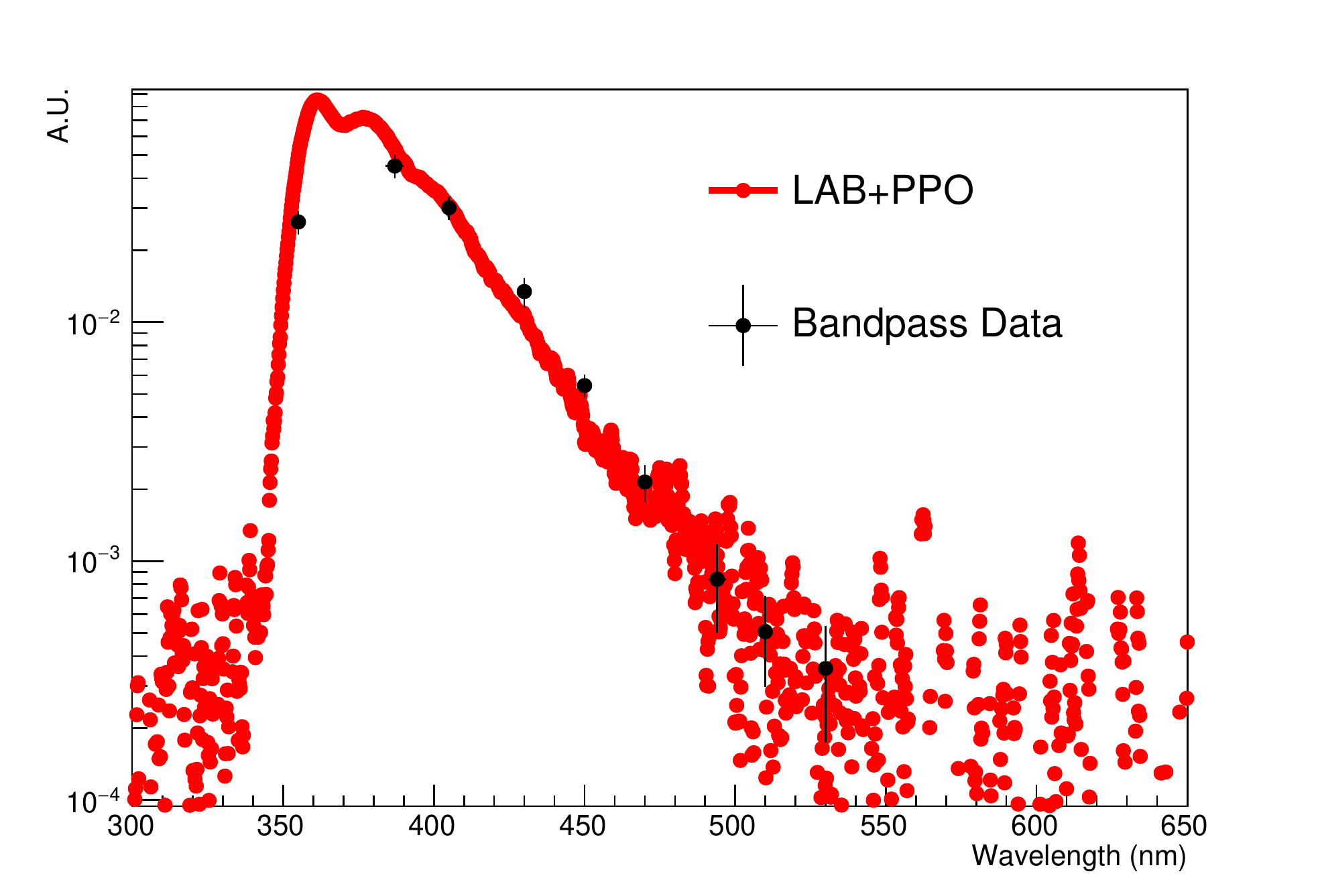}
\caption{The measured emission spectrum compared to the amount of scintillation light detected for each bandpass filter. The amount of scintillation light is scaled based on the total number of triggered events, the efficiency of the PMT, and the transmission of the filter. The shape extracted from the bandpass filter measurement agrees well with the emission spectrum. The noise above 550~nm and below 340~nm is from the dark noise of the spectrometer.}
\label{fig:emisband}
\end{figure}

In the bandpass data the total amount of scintillation light per triggered event is scaled based on the expected transmission of the filter and efficiency of the PMT. The result is shown in Figure \ref{fig:emisband}, compared directly to the measured emission spectrum. The bandpass filter data is consistent with our measured emission spectrum, including the long wavelength emission beyond 450~nm. 

It was initially unclear if some of the Cherenkov light detected in our setup comes from $\beta$s that enter the acrylic cube after depositing some energy in the scintillator. We used a GEANT4-based Monte Carlo to estimate the size of this effect, and have found that removing the acrylic block from the simulation does not change the fraction of Cherenkov light.

\section{Spectral Sorting Using Dichroic Filters}

While bandpass filters allow a good Cherenkov and scintillation separation as shown above, they have the obvious disadvantage that much of the light is lost.  Most photon-based detectors need to optimize the light yield, and while the discrimination between Cherenkov and scintillation light is useful, it nonetheless is not likely to be worth a large-scale loss of light.

Fortunately, as we will show below, dichroic filters provide a way of `sorting' photons by wavelength, without necessarily losing light. Dichroic filters reflect or transmit wavelengths around a specified `cut-on' wavelength, with little absorption.  If the filter is a `longpass' filter and thus transmits long-wavelength photons, the broad-band Cherenkov light is passed along with a very small amount of scintillation light, depending on the value of the cut-on wavelength.  Short-wavelength photons are reflected, and would include the majority of the scintillation photons along with the shorter-wavelength Cherenkov photons.  Separation of the short-wavelength light could be done with fast timing, as has been shown by the CHESS detector~\cite{berkeley}.

\begin{figure}[ht!]
    \centering
    \includegraphics[scale=0.37]{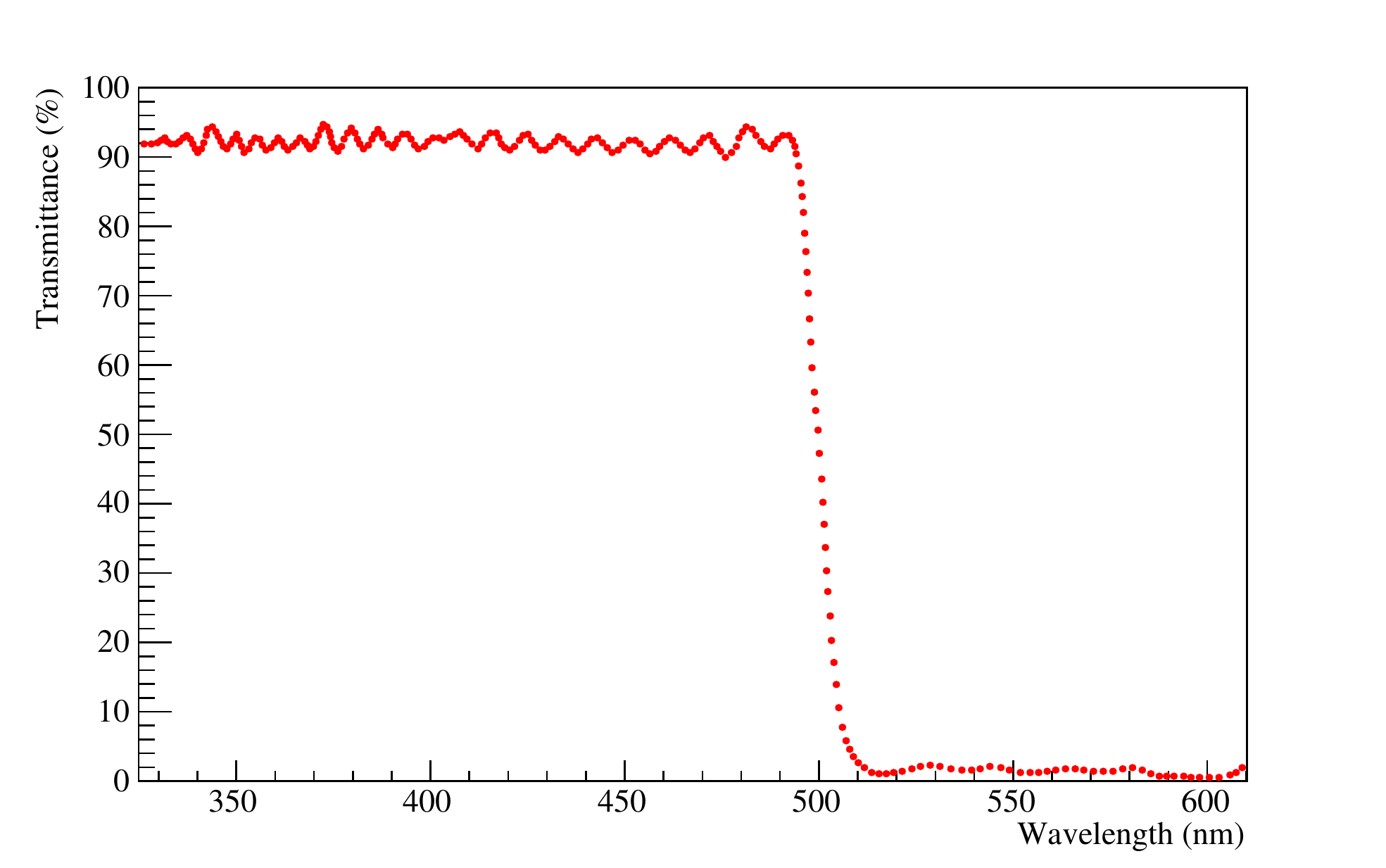}
    \includegraphics[scale=0.37]{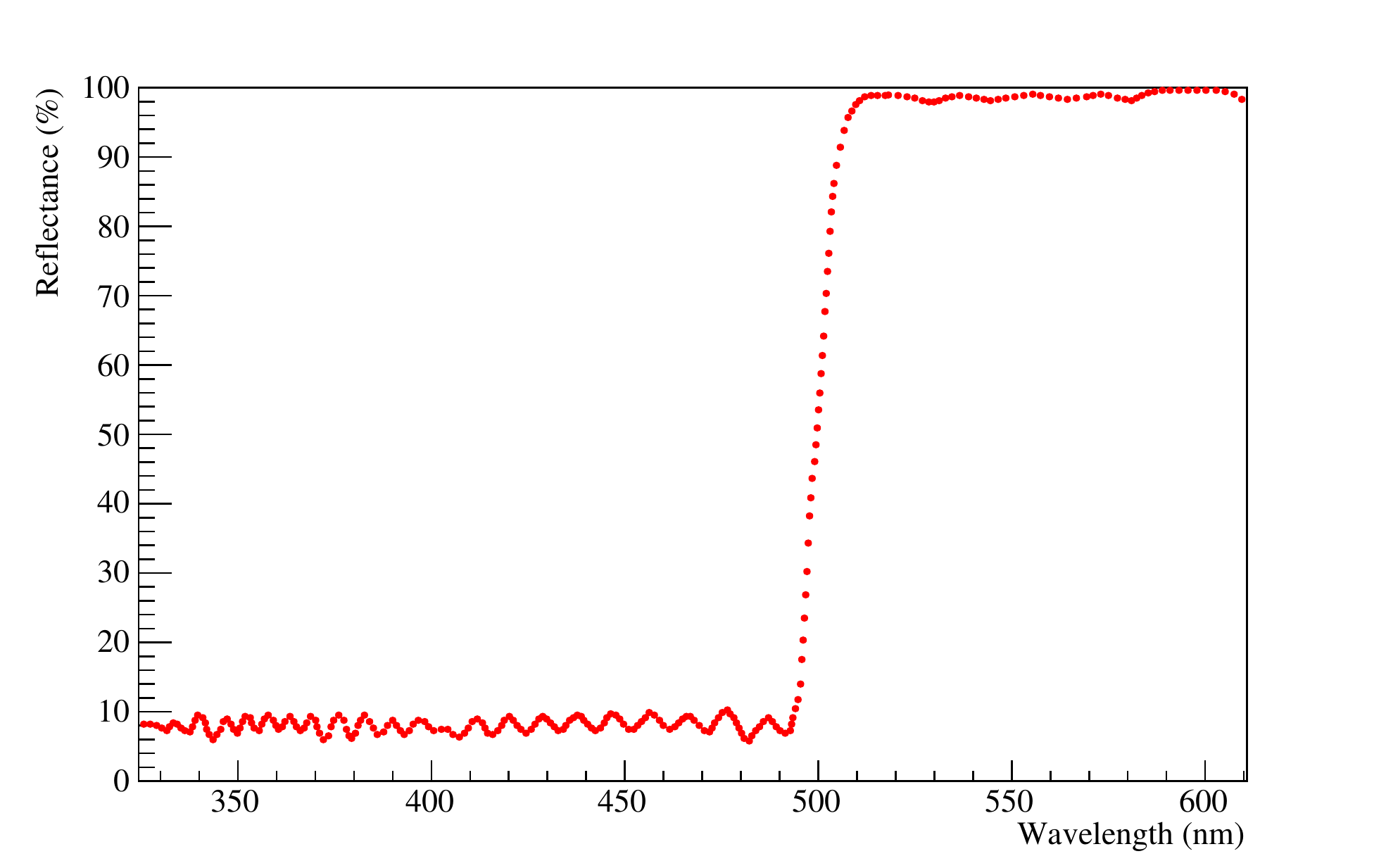}
    \caption{Example transmission and reflection curves for the 500~nm shortpass filter at a 45$^{\circ}$ incident angle, as provided by Edmund Optics \cite{edmunddichroicSP}.}
    \label{fig:transm_refl}
\end{figure}

\begin{figure}[ht!]
    \centering
    \includegraphics[scale=0.37]{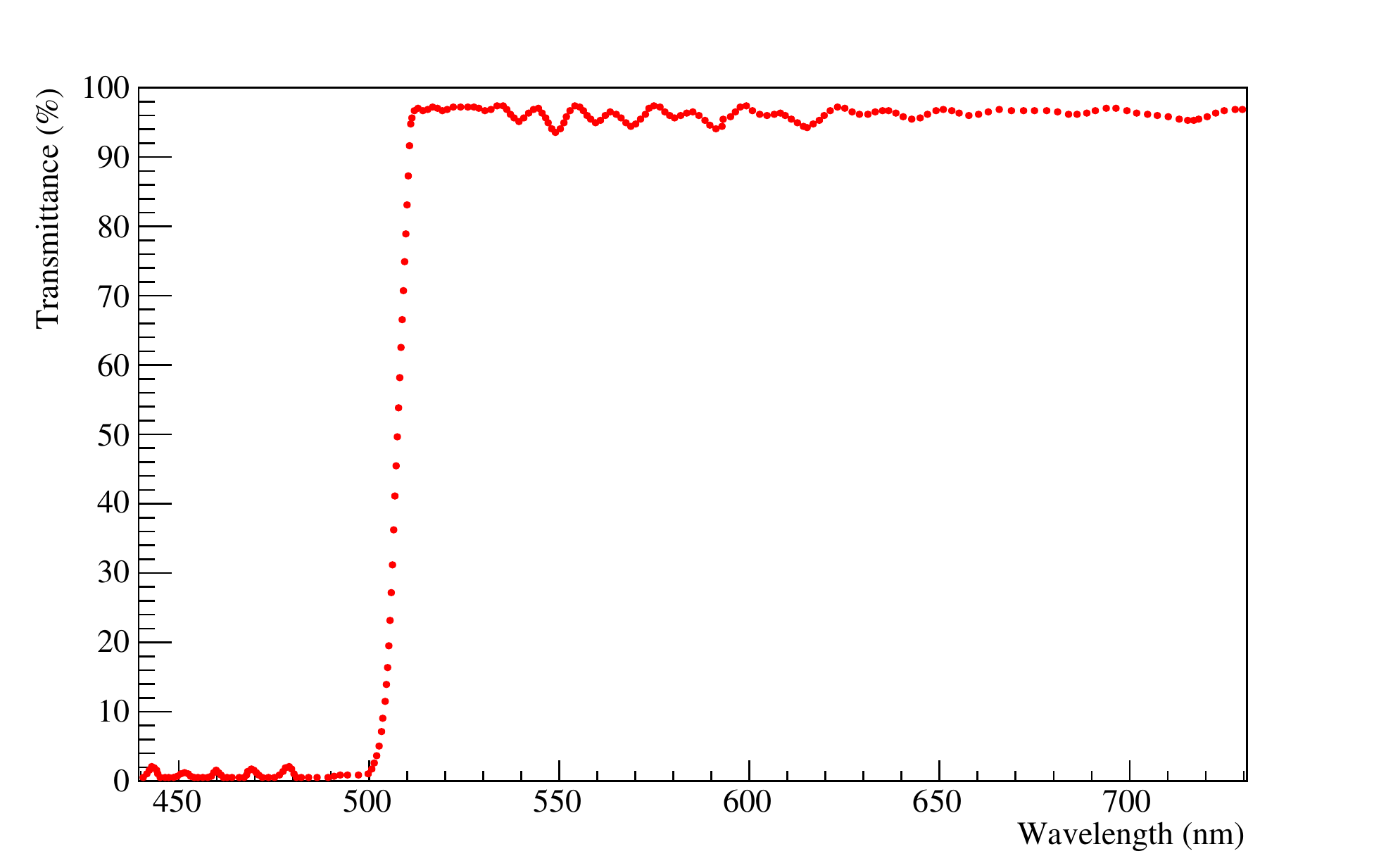}
    \includegraphics[scale=0.37]{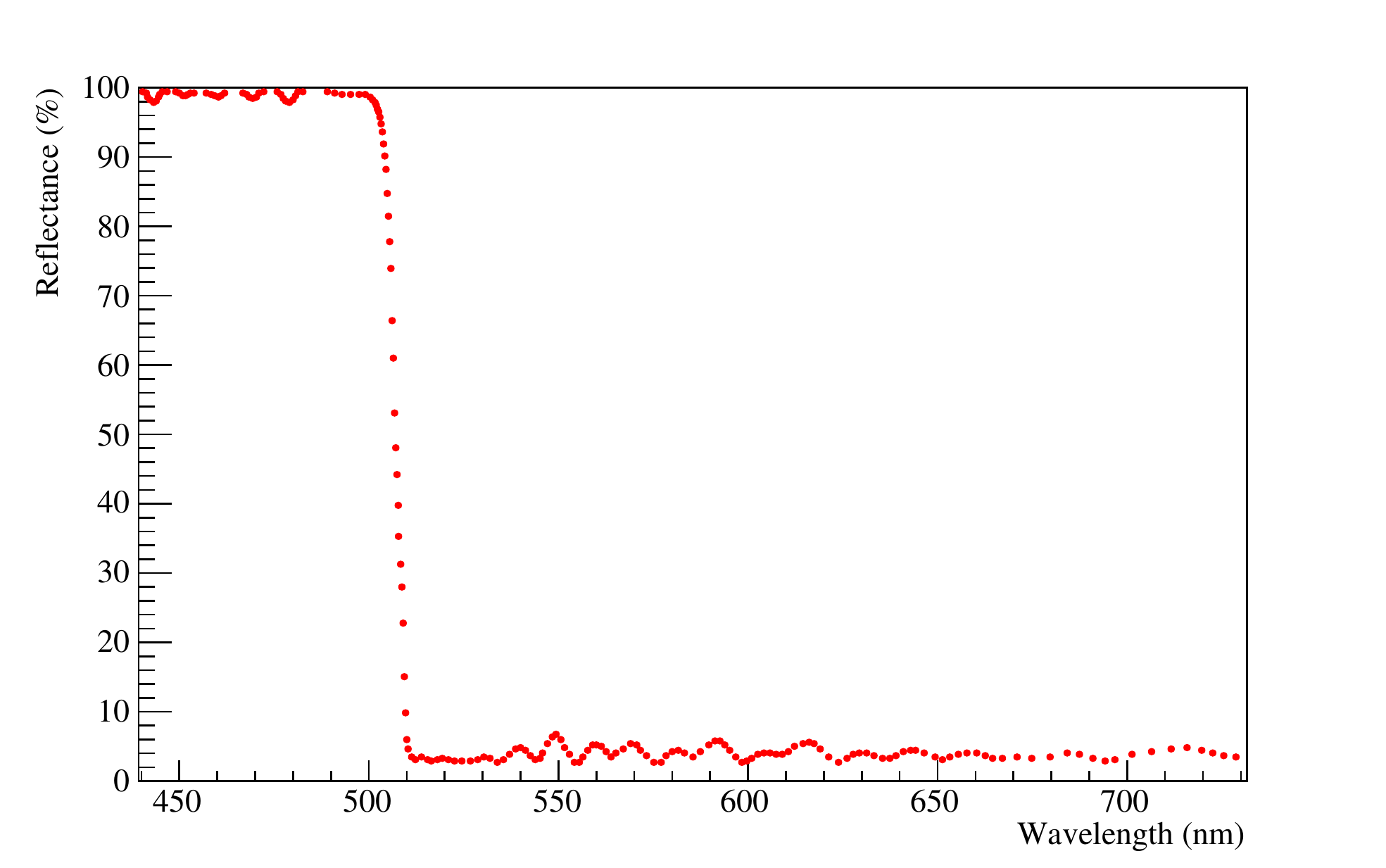}
    \caption{Example transmission and reflection curves for the 506~nm longpass filter at a 45$^{\circ}$ incident angle, as provided by Edmund Optics \cite{edmunddichroicLP}.}
    \label{fig:transm_refl_2}
\end{figure}

Our dichroic filter apparatus was designed to detect both the transmitted and reflected light, and is described below in Section \ref{sec:dichroicexp}. The measurement is performed with two different dichroic filters, which were chosen to roughly complement each other. These filters were purchased from Edmund Optics, who provide lower limits on the specifications of the filters. The longpass filter has a transmittance of $>$90\% for wavelengths longer than 506~nm and a reflectance of $>$98\% for shorter wavelengths. The shortpass filter has a transmittance of $>$85\% for wavelengths shorter than 500~nm and a reflectance of $>$95\% for wavelengths longer than 500~nm. Edmund optics provides example transmission and reflection curves for these filters \cite{edmunddichroicSP, edmunddichroicLP}, which are shown in Figures \ref{fig:transm_refl} and \ref{fig:transm_refl_2}. Note that although the shortpass filter is only guaranteed to transmit $>$85\%, most of the remaining 15\% of the light is reflected rather than absorbed, so it can still be detected. This is similarly true for the longpass filter.

With these filters, we can sort the short-wavelength (primarily scintillation) light toward one PMT and the long wavelength (including broad-band Cherenkov) light primarily to the other. Our results for spectral sorting include light produced in LAB scintillator with either PPO or PTP, the latter providing better separation because of a shorter cut-on wavelength resulting in more Cherenkov light available in the long wavelength bin.

In a practical, large-scale experiment, the sorting would have to be done in a way that ensures as little light loss as possible while still being able to be built inexpensively and simply, likely with large-area photosensors like large-area PMTs.  We include below results on 8'' Hamamatsu PMTs, to show the effects of their relatively poorer timing. In a subsequent publication, we will discuss our design of a dichroic Winston cone that could accomplish the sorting while fitting into large-scale detectors.

\subsection{Experimental Setup}\label{sec:dichroicexp}

\begin{figure}[ht!]
    \centering
    \includegraphics[trim = 2.0cm 9.5cm 0.5cm 9cm, clip=true, scale=0.6]{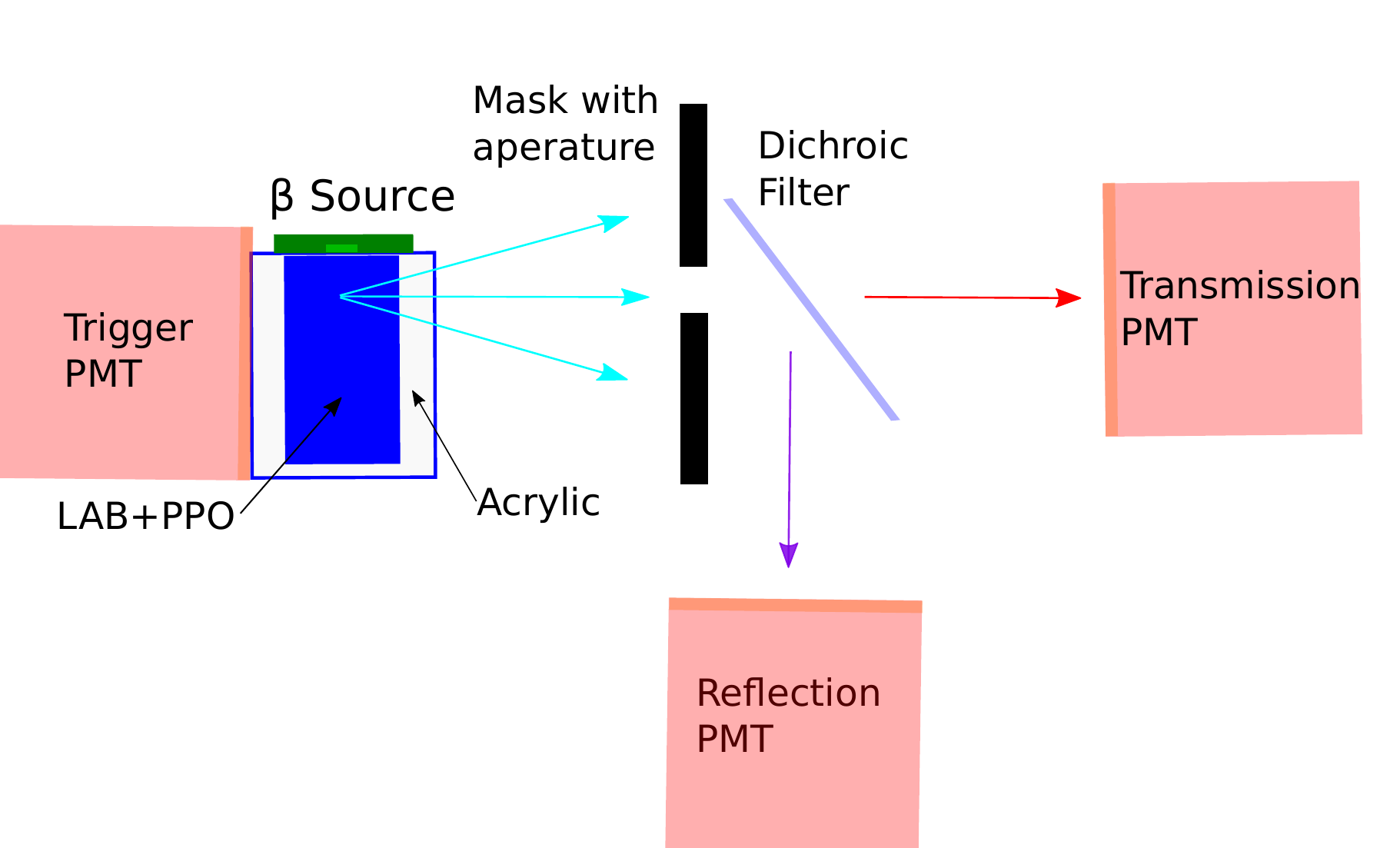}
    \caption{A schematic of the experimental setup with the dichroic filter. The setup is the similar to Figure \ref{fig:schematic}, with the replacement of the bandpass filter with a dichroic filter and the addition of the reflection PMT. Note the measurement is performed with both LAB+PPO as well as LAB+PTP.}
    \label{fig:schematic_dichroic}
\end{figure}

The experimental setup is very similar to the one shown in Figure \ref{fig:schematic}. However, instead of a bandpass filter, a dichroic filter placed at an incident angle of 45$^{\circ}$ is placed behind the aperture. Additionally, a third PMT, called the reflection PMT, is placed to view the reflected light from the dichroic filter. The reflection PMT is the same type as the other two PMTs and is run at the same high voltage. This setup is shown schematically in Figure \ref{fig:schematic_dichroic}. The dichroic filters are both designed for optimal performance at 45$^{\circ}$, so a rotating stage with 1$^{\circ}$ accuracy was used to hold the filter. 

The measurement is performed with PPO as well as another primary fluor called PTP. The PTP is very similar to the PPO and dissolved in the LAB at 2~g/L. Both the emission time constants and light yield of LAB+PTP are similar to LAB+PPO; however, the emission spectrum peaks lower in wavelength, around 340~nm, as shown in Figure \ref{fig:spectra}. This allows one to use a shorter wavelength dichroic filter to achieve the same level of separation, but with a higher total number of Cherenkov photons detected, as is discussed in Section \ref{sec:dichroicresults}.

\subsection{Results}\label{sec:dichroicresults}

The data analysis techniques described in Section \ref{sec:analysis} are identical for this analysis, except for the inclusion of the reflection PMT, which is treated identically in terms of calculating the $\Delta$t from the trigger PMT. Additionally the histograms are fit using the same method as described in Section \ref{sec:results}. 

\begin{figure}[ht!]
\centering
\includegraphics[scale=0.35]{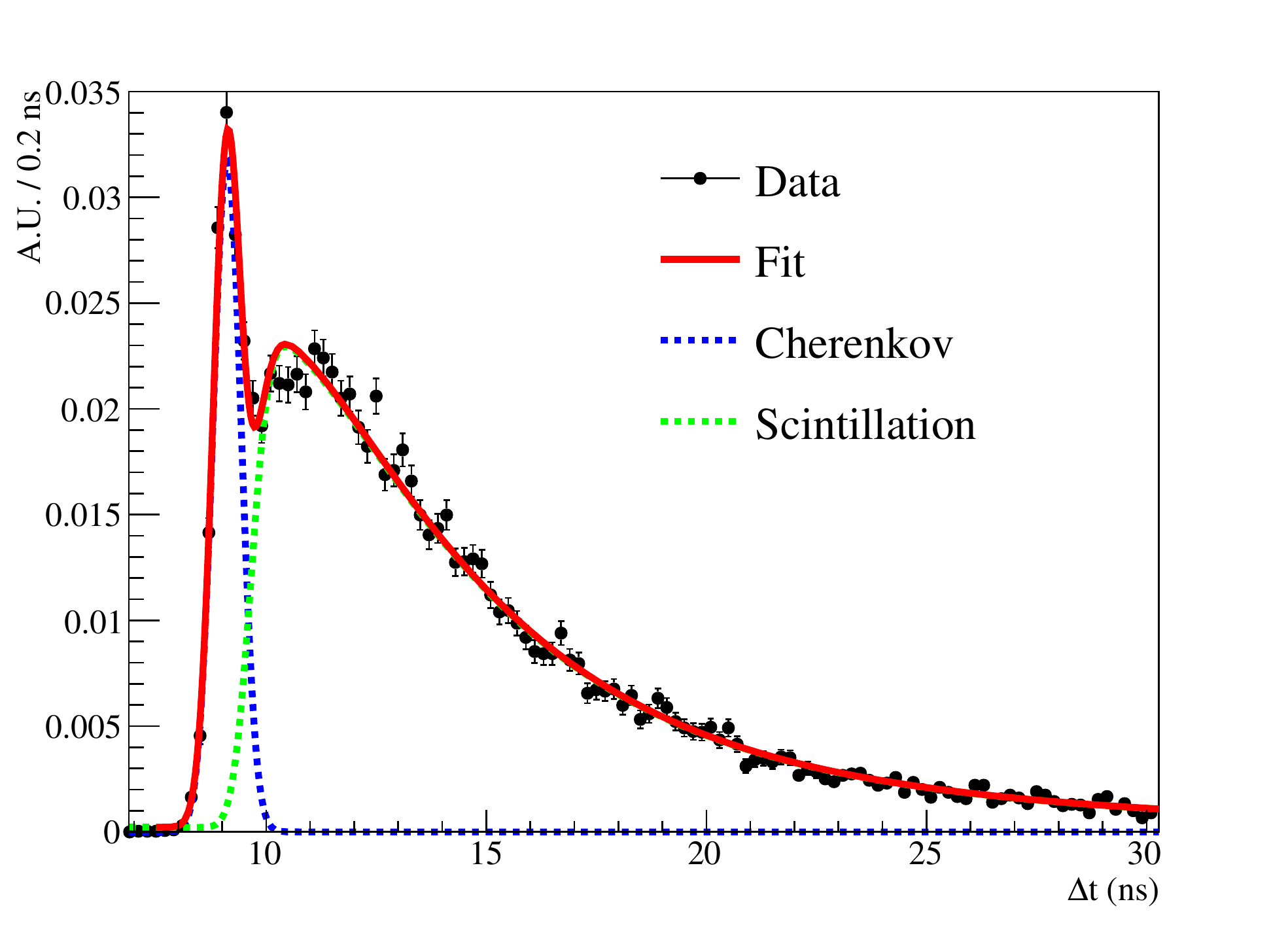}
\includegraphics[scale=0.35]{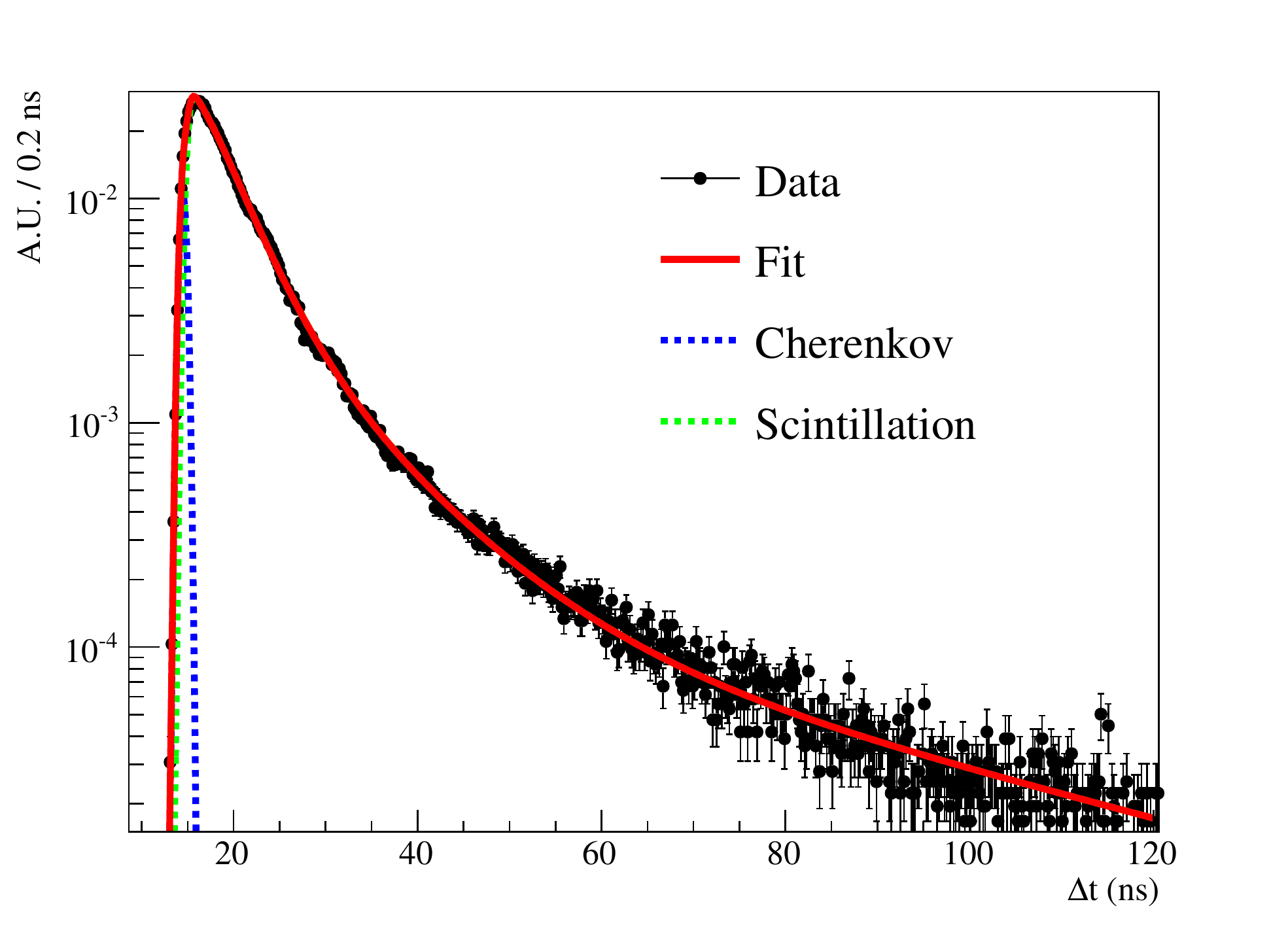}
\caption{The fit results for the longpass 506~nm dichroic filter using LAB+PPO. The transmitted light (left) shows clear Cherenkov and scintillation separation, while the reflected light (right) shows the LAB+PPO emission time-profile.}
\label{fig:dichroic_fits_LP}
\end{figure}

\begin{figure}[ht!]
\centering
\includegraphics[scale=0.35]{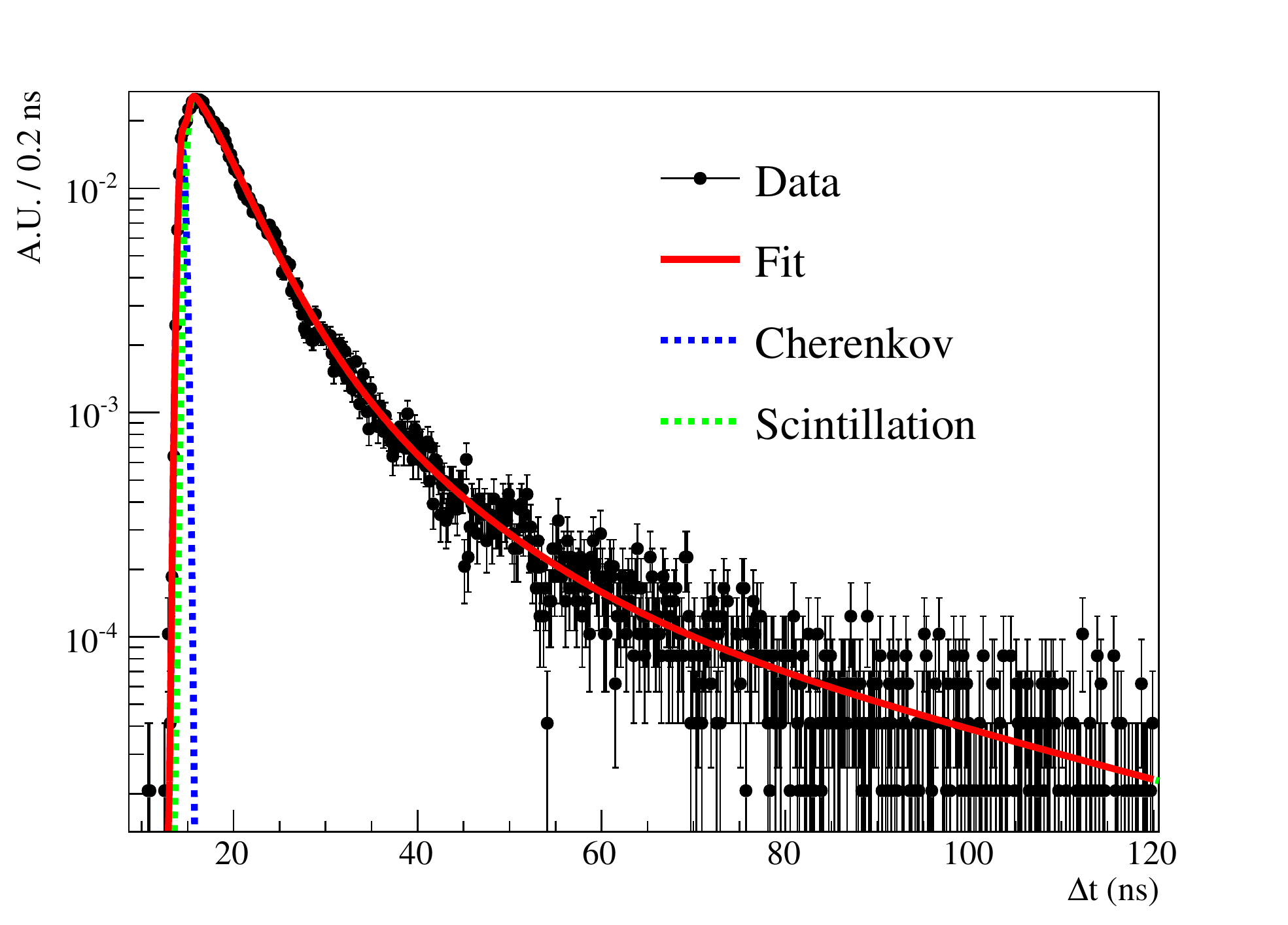}
\includegraphics[scale=0.35]{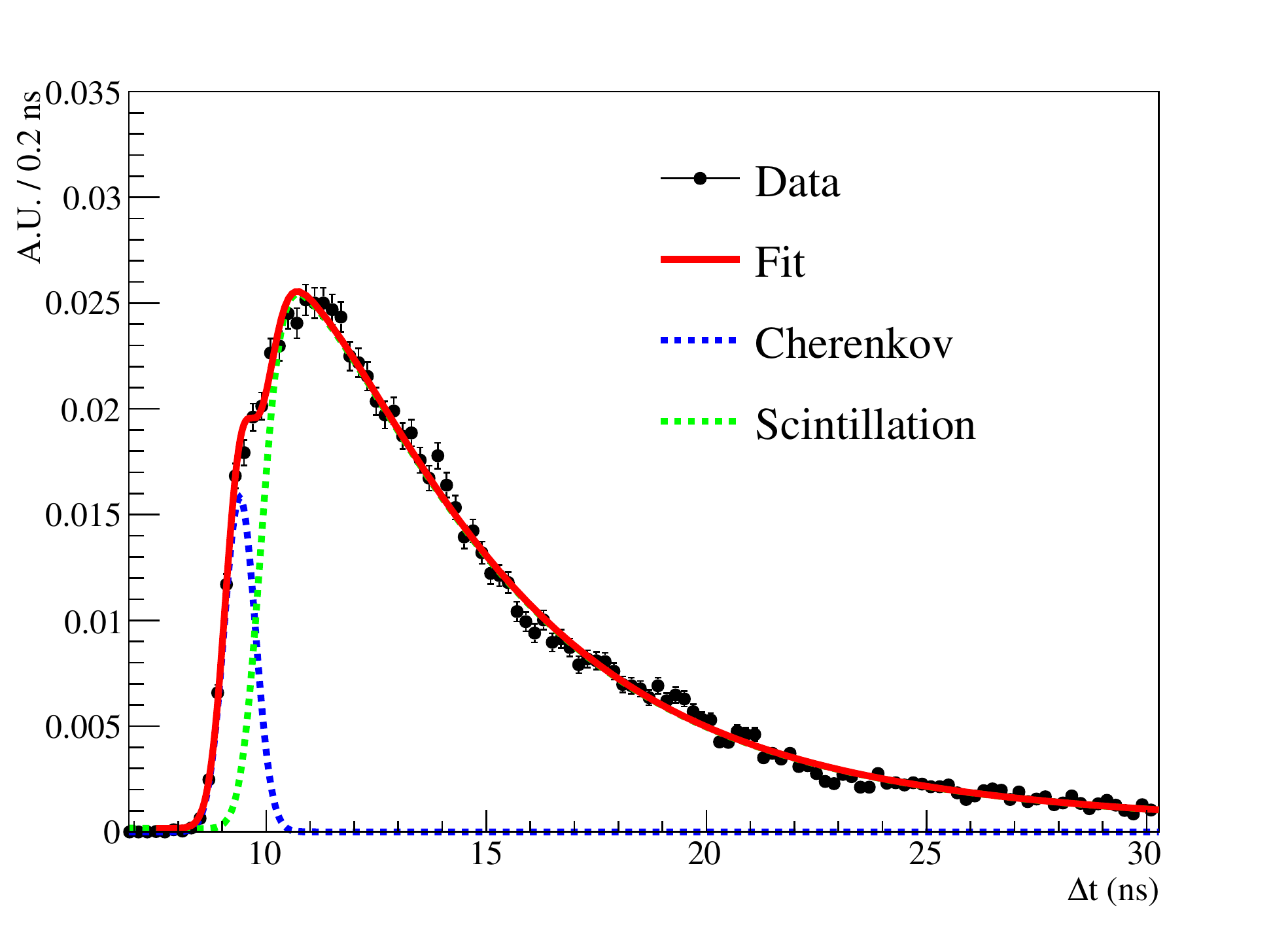}
\caption{The fit results for the shortpass 500~nm dichroic filter using LAB+PPO. The transmitted light (left) shows the LAB+PPO emission time-profile, while the reflected light (right) shows modest Cherenkov and scintillation separation.}
\label{fig:dichroic_fits_SP}
\end{figure}

Shown in Figures \ref{fig:dichroic_fits_LP} and \ref{fig:dichroic_fits_SP} are the fits for the longpass and shortpass filter respectively using LAB+PPO. The data for both the transmission PMT and the reflection PMT are shown with the corresponding fit. For the longpass filter, only the long wavelength light is transmitted, and a clear Cherenkov peak can be identified at the transmission PMT, similar to the bandpass data presented. The reflected light is primarily the LAB+PPO scintillation light, and the data for the reflection PMT data shows the typical LAB+PPO scintillation time-profile. For the 500~nm shortpass filter the reflected light shows more modest Cherenkov separation at prompt times. The shortpass filter's performance was not quite as good (in terms of the separation) as the longpass filter, which is primarily due to the non-negligible reflection at the LAB+PPO emission wavelengths, as can be seen in Figure \ref{fig:transm_refl}.

The fraction of lost photons can be estimated by comparing the sum of the total number of detected photons in the reflection and transmission PMTs to the total number of detected photons in the measurement with no filter (described in Section \ref{sec:results}). This was found to be 97.1 $\pm$ 1.5\% and 98.7 $\pm$ 1.5\% for the longpass and shortpass filters respectively. The uncertainties are primarily systematic and were estimated by swapping the transmission and reflection PMTs and taking the difference as a two-sided uncertainty. 

\begin{figure}[ht!]
\centering
\includegraphics[scale=0.35]{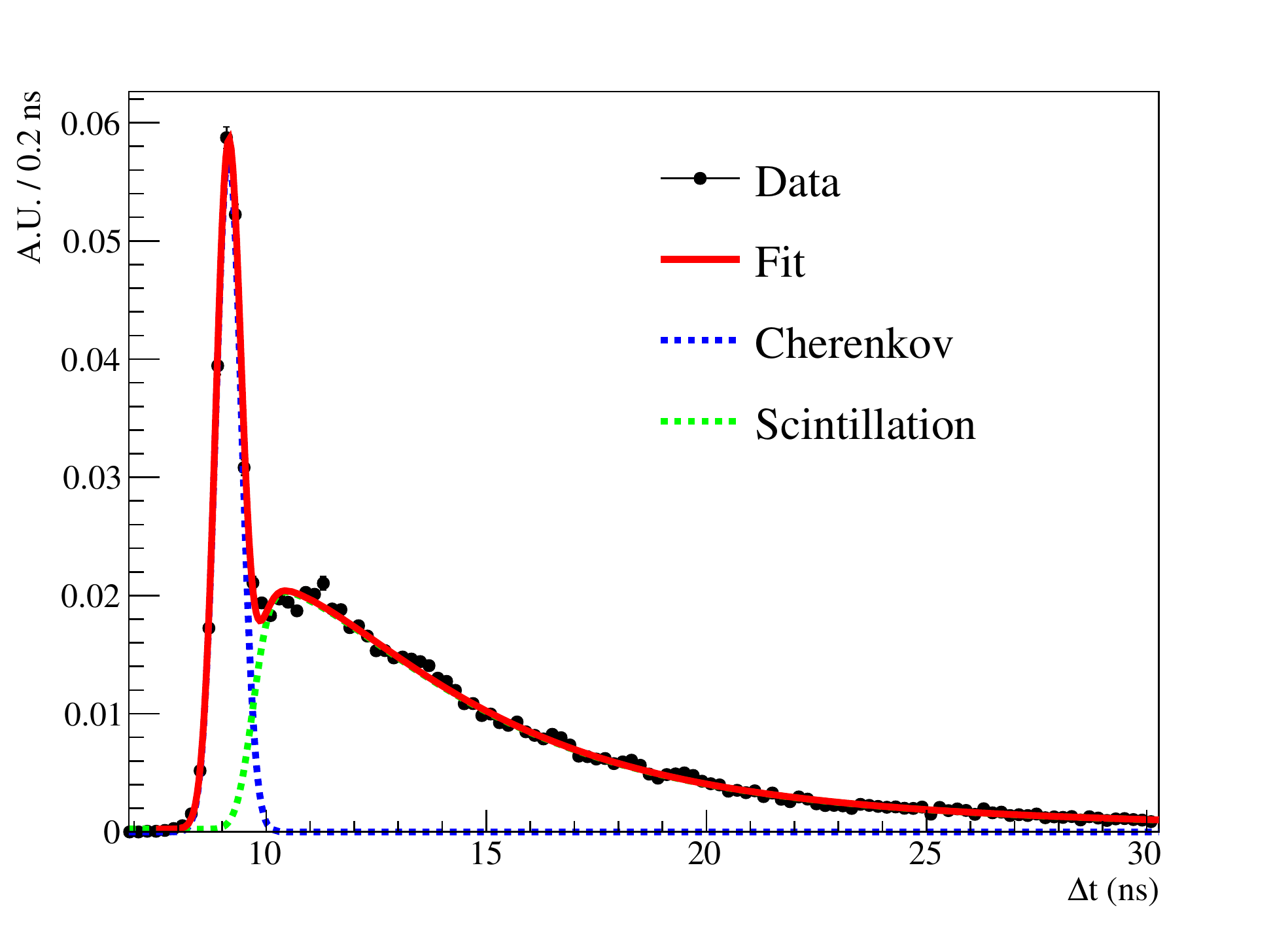}
\includegraphics[scale=0.35]{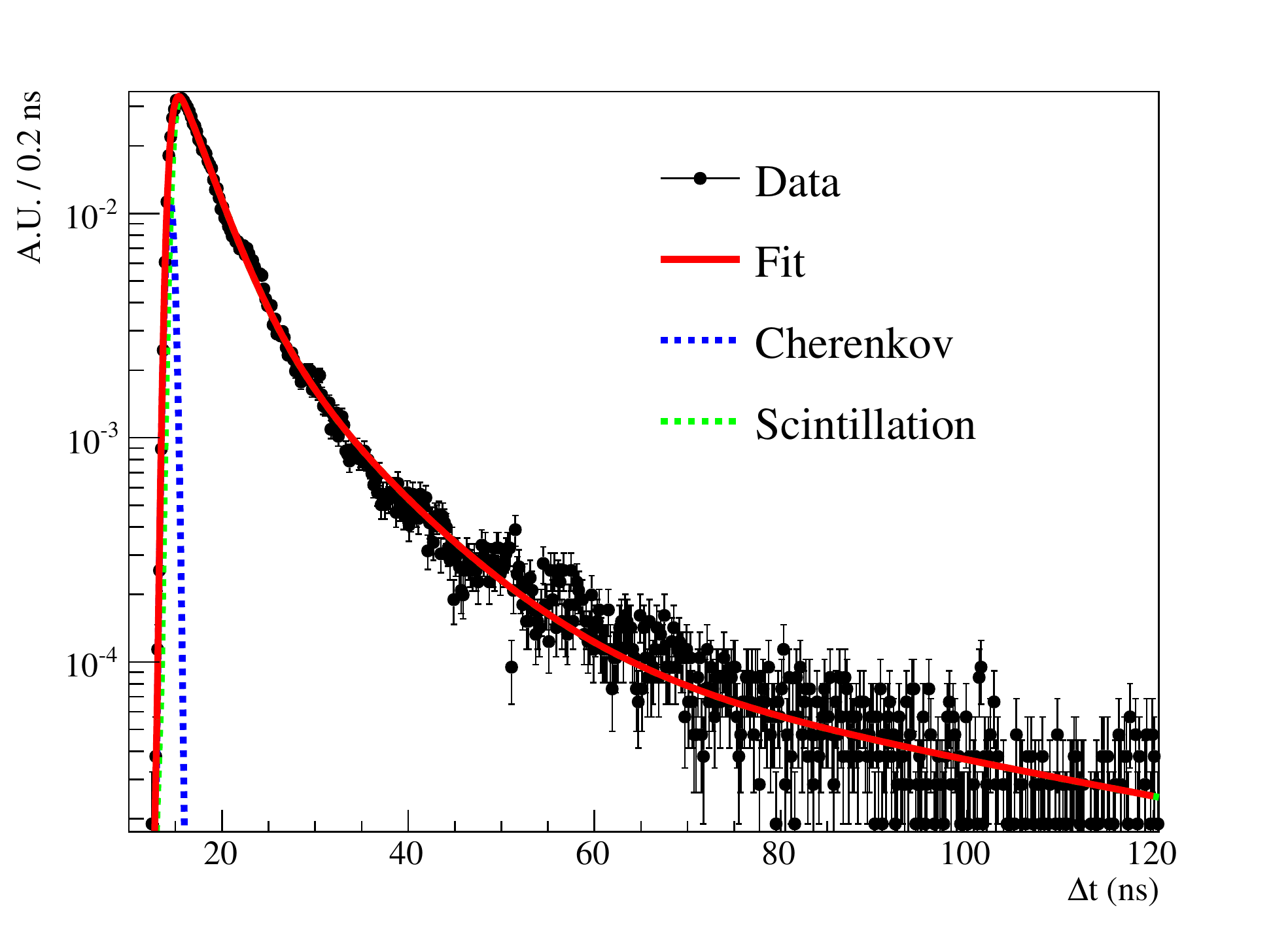}
\caption{The fit results for the longpass 450~nm dichroic filter using LAB+PTP. The transmitted light (left) shows clear Cherenkov and scintillation separation, while the reflected light (right) shows the LAB+PTP emission time-profile.}
\label{fig:dichroic_fits_PTP_LP}
\end{figure}

Figure \ref{fig:dichroic_fits_PTP_LP} shows the results for a 450~nm longpass filter using LAB+PTP. Because the emission spectrum is shorter in wavelength than the PPO, the PTP allows one to use a shorter wavelength dichroic filter while maintaining good separation. This increases the total amount of Cherenkov light detected due to the larger acceptance for the shorter wavelength Cherenkov light, in addition to the higher quantum efficiency of the PMT at shorter wavelengths. 

\subsection{Results with a Large Area PMT} 

\begin{figure}[ht!]
\centering 
\includegraphics[scale=0.6]{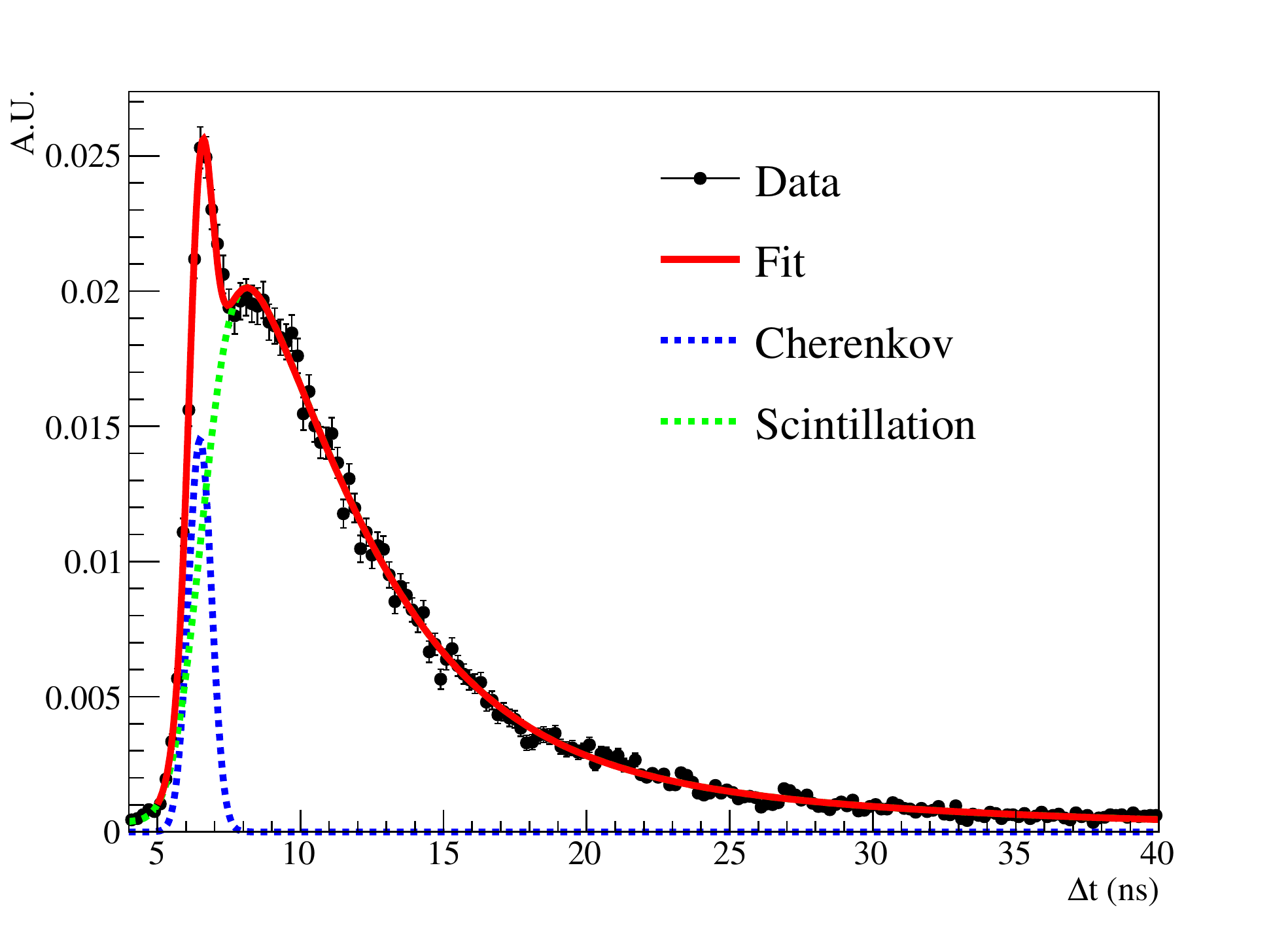}
\caption{The data using the R5912-MOD PMT to detect the transmitted light through the longpass dichroic filter. The fit result shows a clear Cherenkov peak at early times.}
\label{fig:dichroicr5912}
\end{figure}

The PMTs used in the measurement presented thus far have extremely good timing properties, and are probably unrealistically expensive and small for a large-scale detector. The transit-time spread of modern large area tubes has been improving, however. Recently Hamamtasu has developed a prototype version of the R5912 8-inch PMT with a TTS of around 700~ps \cite{r5912mod}, called the R5912-MOD. The PMT was deployed in an identical setup to Figure \ref{fig:schematic_dichroic}, with the transmission PMT replaced with the R5912-MOD PMT. For this measurement we used the 506~nm dichroic longpass filter in order to identify the long-wavelength Cherenkov light. The results of this measurement are shown in Figure \ref{fig:dichroicr5912} with the corresponding fit. The only change to the fit is to allow for a wider PMT TTS, which fit to 820 $\pm$ 150~ps, consistent with the results in \cite{r5912mod}. The fit parameters are consistent with those presented in Table \ref{tab:fitresults}. With an adjusted prompt window, a purity, $P$, of 52 $\pm$ 4 \% is found. The lower purity is expected due to the broadening of the PMT TTS. 

\section{Conclusion}

Using bandpass filters to look at specific segments of the LAB+PPO emission spectrum has demonstrated separation of Cherenkov and scintillation light. In particular, in the tail of the PPO emission spectrum above 450~nm the Cherenkov light is clearly temporally separated from the scintillation light. A fit to the time profiles that includes components for both the fast Cherenkov light and the slower scintillation light indicates that a very pure selection of Cherenkov light can be selected using a simple prompt time cut. Additionally the amount of scintillation light extracted from the fits is consistent with the expected emission spectrum. 

Our measurements with dichroic filters show the ability to `sort' photons towards two different PMTs, one detecting the reflected light and the other detecting the transmitted light. With this approach, we can separate the Cherenkov and scintillation components while still detecting a large fraction of the scintillation light. The method works well even  with a fast large-area PMT, making it interesting for future large-scale scintillator experiments which hope to maintain a high light yield, while also achieve scintillation and Cherenkov separation.

\section{Acknowledgements}
Thanks to Gabriel Orebi Gann for helpful discussion about Cherenkov/scintillation separation and future application with THEIA. Thanks to Gene Beier, Ian Coulter, Eric Marzec, Logan Lebanowski, and Nuno Barros for useful discussion regarding the LAB+PPO timing and emission spectra. Thanks to Tony LaTorre for developing the Lecrunch software used for the data readout and for suggestions regarding fitting the timing spectrum. Thanks to an anonymous reviewer for suggesting measurements that resulted in the correction of several mistakes. This work was supported by the Department of Energy and the Office of High Energy Physics, under grant number DE-FG02-88ER40479.


\clearpage

\begin{thebibliography}{12}
\bibitem{snoplus} SNO+ Collaboration (S. Andringa et al.), \textit{Current Status and Future Prospects of the SNO+ Experiment}, Advances in High Energy Physics, vol. 2016, 6194250
\bibitem{juno} JUNO Collaboration, (F. An et al.), \textit{Neutrino Physics with JUNO}, J. Phys. G 43 (2016) 030401
\bibitem{dayabay2} W. Beriguete. et al., \textit{Production of Gadolinium-loaded Liquid Scintillator for the Daya Bay Reactor Neutrino Experiment}, arXiv:1402.6694 \textbf{[physics.ins-det]}
\bibitem{theia} G. D. Orebi Gann, \textit{Physics Potential of an Advanced Scintillation Detector: Introducing THEIA}, arXiv:1504.08284 \textbf{[physics.ins-det]}
\bibitem{rbonvgdog} R. Bonventre, G. D. Orebi Gann, \textit{Sensitivity of a low threshold directional detector to CNO-cycle solar neutrinos}, Eur. Phys. J. C (2018) 78: 435
\bibitem{biller} S. Biller, \textit{Probing Majorana neutrinos in the regime of the normal mass hierarchy}, Physical Review D, 071301R, 8 April 2013
\bibitem{dayabay} Li Xiao-Bo et al., \textit{Timing properties and pulse shape discrimination of LAB-based liquid scintillator}, Chinese Physics C. 35. 1026. 10.1088/1674-1137/35/11/009
\bibitem{queens} O'Keefe et al., \textit{Scintillation decay time and pulse shape discrimination in oxygenated and deoxygenated solutions of linear alkylbenzene for the SNO+ experiment}, Nucl. Instrum. Meth. A 640:119-122, 2011
\bibitem{lena} T. Marrodan Undagoitia et al., \textit{Fluorescence decay-time constants in organic liquid scintillators}, Rev. Sci. Instrum. 80:043301, 2009
\bibitem{ranucci} Lombardi et al., \textit{Decay time and pulse shape discrimination of liquid scintillators based
on novel solvents}, Nucl. Instrum. Meth. A 701:133-144, 2013
\bibitem{chsp} C. Aberle et al., \textit{Measuring Directionality in Double-Beta Decay and Neutrino Interactions with Kiloton-Scale Scintillation Detectors}, arXiv:1307.5813 \textbf{[physics.ins-det]}
\bibitem{bh} M. Li et al., \textit{Separation of Scintillation and Cherenkov Lights in Linear Alkyl Benzene}, Nucl. Instrum. Methods Phys. Res. A 830 (2016) 303-308
\bibitem{berkeley} J Caravaca et al., \textit{Cherenkov and scintillation light separation in organic liquid
scintillators}, Eur. Phys. J. C (2017) 77: 811
\bibitem{spc} L.M. Bollinger et al., \textit{Measurement of the Time Dependence of Scintillation Intensity by a Delayed Coincidence Method}, Review of Scientific Instruments 32, 1044 (1961)
\bibitem{photochem} Taniguchi, M.; Lindsey, J. S.  \textit{Database of Absorption and Fluorescence Spectra of >300 Common Compounds for Use in PhotochemCAD}, Photochem. Photobiol. 2018, 94, 290-327
\bibitem{hamamatsu} Hamamatsu Photonics Datasheet, Viewed Nov. 19 2018, \\
https://www.hamamatsu.com/resources/pdf/etd/R7600U\_TPMH1317E.pdf
\bibitem{lecrunch} A. Latorre, \textit{Lecrunch Open Source Software}, Viewed Nov. 19 2018, \\ https://bitbucket.org/tlatorre/lecrunch/
\bibitem{selfabsorption} T. M. Undagoitia, \textit{Measurement of light emission in organic liquid scintillators and studies towards the search for proton decay in the futrue large scale detector LENA}, PhD thesis, Technical University of Munich, Viewed Nov. 19, 2018
\bibitem{thorlabs} Thorlabs, Transmission Of Bandpass Filters, Viewed Nov. 19, 2018, \\
https://www.thorlabs.com/newgrouppage9.cfm?objectgroup\_id=1860\&pn=FLH355-10
\bibitem{edmund} Edmund Optics, Transmission Of Bandpass Filters, Viewed Nov. 19, 2018,  \\
https://www.edmundoptics.com/p/387nm-cwl-25mm-dia-11nm-bandwidth-od-6-fluorescence-filter/27202/
\bibitem{etel} N. Barros et al., \textit{Characterization of the ETEL D784UKFLB 11 in. photomultiplier tube}, Nucl. Instr. Meth. A852 (2017) 15
\bibitem{roofit} W. Verkerke, D. Kirkby, \textit{The RooFit toolkit for data modeling}, arXiv:physics/0306116 \textbf{[physics.data-an]}
\bibitem{ly} J.B. Cummings et al., \textit{Improving Light Yield Measurements for Low-Yield Scintillators}, arXiv:1810.02885 \textbf{[physics.ins-det]}
\bibitem{emission} X. Hua-Lin et al., \textit{Study of absorption and re-emission processes in a ternary liquid scintillation system}, Chinese Physics C, Volume 34, Issue 11, pp. 1724-1728 (2010).
\bibitem{edmunddichroicSP} Edmund Optics, \textit{500nm Dichroic Shortpass Filter}, Viewed Nov. 19, 2018,  \\
https://www.edmundoptics.com/p/500nm-12.5-x-17.6mm-dichroic-shortpass-filter-/23532/
\bibitem{edmunddichroicLP} Edmund Optics, \textit{506nm Dichroic Longpass Filter}, Viewed Nov. 19, 2018,  \\
https://www.edmundoptics.com/p/506nm-25.2-x-35.6mm-high-performance-fluorescence-dichroic-filter/3970/
\bibitem{hyperk} C. Rott, S. In, F. Retiere and P. Gumplinger, \textit{Enhanced Photon Traps for Hyper-Kamiokande}, JINST 12, no. 11, P11021 (2017) arXiv:1708.01702 [astro-ph.IM]
\bibitem{arapuca} E. Segreto et al., \textit{Liquid Argon test of the ARAPUCA device}, arXiv:1805.00382 \textbf{[physics.ins-det]}
\bibitem{r5912mod} T. Kaptanoglu, \textit{Characterization of the Hamamatsu 8'' R5912-MOD Photomultiplier tube}, 
Nucl.\ Instrum.\ Meth.\ A {\bf 889}, 69 (2018) doi:10.1016/j.nima.2018.01.086.
\end{thebibliography}
\end{document}